\documentclass[11pt]{article}

\usepackage{authblk}

\usepackage{amsmath,amssymb,amsthm}
\usepackage{amsfonts}            
\usepackage{mathrsfs}            
\usepackage{mathdots}
\usepackage{extarrows}
\usepackage{multirow}
\usepackage{multicol}
\usepackage{makecell}
\usepackage{enumerate}
\usepackage{graphicx,psfrag,epsf}
\graphicspath{{figs/}}
\usepackage{ulem} 
\usepackage{url}
\usepackage{color,subcaption}
\usepackage{xcolor}
\usepackage{geometry}
\usepackage{algorithm}
\usepackage{algpseudocode}
\usepackage{caption}
\usepackage[counterclockwise]{rotating}
\usepackage[T1]{fontenc}
\usepackage{setspace}
\usepackage{bm} 

\usepackage[export]{adjustbox}
\usepackage{wrapfig} 

\usepackage{thmtools}
\usepackage{thm-restate}
\usepackage[colorlinks,linkcolor=blue,anchorcolor=blue,citecolor=blue]{hyperref}

\usepackage{booktabs}
\usepackage{footnote}
\usepackage[flushleft]{threeparttable}
\usepackage{longtable}
\usepackage{verbatim}

\usepackage{listings} 

\usepackage{float}

\usepackage[round]{natbib}
\usepackage{cite}

\hypersetup{pdfborder = {0 0 0},colorlinks=true,linkcolor=blue,citecolor=blue}
\newcommand{\indep}{\rotatebox[origin=c]{270}{$\models$}}
\newcommand{\tabincell}[2]{\begin{tabular}{@{}#1@{}}#2\end{tabular}}

\newtheorem{thm}{Theorem}[section]

\newtheorem{pro}{Proposition}

\makeatletter
\@addtoreset{equation}{section}
\makeatother

\title{Kernel-Distance-Based Covariate Balancing}

\author{Xialing Wen, Ying Yan\thanks{The Corresponding Author. Email: \href{mailto: yanying7@mail.sysu.edu.cn}{yanying7@mail.sysu.edu.cn}}, Wenliang Pan, Xianyang Zhang \\
School of Mathematics, Sun Yat-sen University,\\ No.135, Xingang West Road, Guangzhou, P.R.China}

\date{}

\hypersetup{final}
\begin{document}

\maketitle

\doublespacing

\begin{abstract}
	A common concern in observational studies focuses on properly evaluating the causal effect, which usually refers to the average treatment effect or the average treatment effect on the treated. In this paper, we propose a data preprocessing method, the Kernel-distance-based covariate balancing, for observational studies with binary treatments. This proposed method yields a set of unit weights for the treatment and control groups, respectively, such that the reweighted covariate distributions can satisfy a set of pre-specified balance conditions. This preprocessing methodology can effectively reduce confounding bias of subsequent estimation of causal effects. We demonstrate the implementation and performance of Kernel-distance-based covariate balancing with Monte Carlo simulation experiments and a real data analysis.
	
	~
\end{abstract}

\section{Introduction}\label{sec:intro}
In causal inference, there are two crucial concepts.
 (i) The propensity score \citep{rosenbaum1983central}: the probability of assignment to the treatment group given the observed covariates.
(ii) The potential outcomes \citep{neyman1923application, rubin1974estimating}: the possible outcomes under different treatment decisions or interventions. The potential outcomes framework, also known as the Neyman-Rubin Potential Outcomes, was first proposed in completely randomized experiments by \citet{neyman1923application} and was extended to both experimental and observational studies by \citet{rubin2005causal}. An overview for the implementation of the potential outcomes framework can be found in \citet{imbens2015causal}. 
Both the two concepts are related to evaluate the causal effect, a central topic in causan inference. Based on various purpose, researchers may prefer estimating the average treatment effect (ATE), the average treatment effect on the treated (ATT), the average treatment effect on the control (ATC), or all of them. 

In observational studies, however, it is not valid to evaluate the causal effect by merely computing the mean-difference in response between different experimental groups due to the confounding variables and the non-random assignment mechanism. One method proposed by \citet{rubin2007design} to solve this problem is to balance the distributions between the treated and the control groups. Two popular preprocessing methods, matching and propensity score methods, have been widely used in observational studies. In causal inference, weighting methods that adjust for observed covariates enjoy substantial popularity. In this paper, we focus on weighting methods.

The weights derived from adjusting the empirical distributions of the observed covariates will be applied to the outcomes such that the reweighted data appear randomized, which can yield stable estimates for the parameters of interest when there is no unmeasured confounders. Generally speaking, inverse probability weighting is less robust and works no better than moment balancing methods. Therefore, we propose a weighting framework, namely the kernel distance-based (KDB) covariate balance, and use the first-moment balance of all covariates as a constraint for the optimization problem. In this article, the KDB method is a preprocessing technique to achieve covariate balance in observational studies with a binary treatment. We give estimators of ATE and ATT based on the KDB framework, among which our primary focus is ATE estimation. Within the same framework, the optimization problem can be flexibly changed to estimate ATT or ATC. 

Subsequently, we further explore their asymptotic properties and compare our proposed method with other preprocessing methods, such as the inverse probability weighting (IPW) \citep{horvitz1952generalization,hirano2003efficient}, the covariate balancing propensity score (CBPS) \citep{imai2014covariate}, the entropy balancing (EB) \citep{hainmueller2012entropy}, and non-parametric calibration weighting (CAL)\citep{chan2016globally}. It can be shown that this proposed framework has several attractive features. Most importantly, the KDB method can automatically match the balance involve the second and possibly higher moments of the covariate distributions as well as interactions. Besides, the KDB estimator for ATE shows greater stability in nonlinear simulation scenarios compared to existing weighting estimators.

The remainder of this article is organized as follows. Section 2 introduces the definitions, notation, and assumptions used throughout the article. Section 3 reviews the main results of four weighting methods, namely the IPW, CBPS, EB, and CAL. Section 4 introduces the weighted kernel distance in detail and describes the theoretical framework of the KDB that we proposed. Section 5 illustrates the proposed method and compares its performance with other weighting methods in two Monte Carlo simulations and a real data analysis. Section 6 concludes this article and discusses further explorations.

\section{Notation and Model Setup}\label{sec:nnms}

\subsection{Notation}\label{subsec:notation}
We use the symbol `$\stackrel{\triangle}{=}$' to distinguish definitions from equalities. 
Suppose the observed dataset consists of a random sample or finite population of $N$ units, each assigned to one of the groups, treatment or control, for which covariate-balanced comparisons are of interest. For each unit $i~(i = 1,\cdots, N)$, we can observe $\{ (Y_i, T_i, \bm{X}_i):~i=1,\cdots,N \}$ of $\{ (Y, T, \bm{X})\}$, where $Y$ is an outcome variable, $T$ is a treatment indicator and $\bm{X} = (X_{1}, \cdots , X_{D})$ with expectation $\bm{\mu}$ and covariance matrix $\bm{\Sigma}$ is a set of $ D $ covariates. Every $T_i$ takes values 1 or 0, which correspondingly indicates the unit $i$ is assigned to the treatment or the control group. We define the sample size in the treatment and control group respectively as $n_1$ and $n_0$. The observed data in the treatment and control groups are denoted as $\{ (Y_{1i}, T_{1i}, \bm{X}_{1i}):~i=1,\cdots,n_1 \}$ and $\{ (Y_{0j}, T_{0j}, \bm{X}_{0j}):~j=1,\cdots,n_0 \}$, respectively, where
       \[ 
       \bm{X}_{1i} = (X_{1i,1}, \cdots , X_{1i,D}) \quad
       , \quad  
       \bm{X}_{0j} = (X_{0j,1}, \cdots , X_{0j,D}).
       \]
The probability of assignment to the treatment group given the
covariates is the so-called propensity score \citep{rosenbaum1983central} 
       \[
        e(\bm{X}) = Pr(T = 1|\bm{X}),
        \]
which plays a central role in causal inference.

Under the potential outcome framework \citep{neyman1923application,rubin1974estimating}, we define a pair of potential outcomes $\{ Y_i(0),Y_i(1) \}$ for every unit $ i $ under treatment 0 or 1 respectively. The observed oucome can be expressed by the pair of potential outcomes under the consistency assumption:
       \[
        Y_i = (1-T_i)Y_i(0) + T_i Y_i(1).
       \]

In this paper, we focus on two causal estimands: the average treatment effect (ATE), that is 
       \[
        \tau := E[Y(1)- Y(0)] = \mu_1 - \mu_0,
       \]
where $\mu_t := E[Y(T=t)]$, and the average treatment effect on the treated (ATT), that is 
       \[
        \tau_1 := E[Y(1)- Y(0) | T = 1] = \mu_{1|1} - \mu_{0|1},
       \]
where $\mu_{t|1} \equiv E[Y(T=t) | T =1]$ for $t=0,1$. The expectations for the potential outcomes given the covariates are
       \[
        \mu_{0}(\bm{X}) \equiv E[Y(0)| X] \quad and  \quad \mu_{1}(\bm{X}) \equiv E[Y(1)| X].
       \]
        
For causal comparisons, we make the following assumptions \citep{rosenbaum1983central} throughout this article so that unbiased estimators of ATE and ATT are available in observational studies.

\subsection{Assumptions}\label{subsec:ass}
\paragraph{Assumption 1 (Strong Ignorability):} \label{para:ass1}~
\\
       \[
       \{Y(0), Y(1)\} \indep T ~|~ \bm{X},
       \]
where $\indep$ denotes independence.

We assume the potential outcomes $ \{Y(0), Y(1) \} $ are independent of the treatment indicator $ T $ given covariates $ \bm{X} $, which implies there is no unmeasured confounders that may cause the selection bias. Formally, as shown by \citet{rosenbaum1983central}, \textbf{Assumption 1} also implies 
       \[
        \{Y(0),Y(1)\} \indep T~ |~ e(\bm{X}),
       \]
and 
       \[
        E(Y | T = 0, \bm{X}) = E(Y | T = 1, \bm{X} ) = E(Y | \bm{X}).
       \]
Under \textbf{Assumption 1}, we can investigate ATE through the value of 
       \[
        \begin{split}
        \tau(\bm{X}) &\equiv  \mu_{1}(\bm{X}) - \mu_{0}(\bm{X}) \\
        &\xlongequal[]{ \{ Y(0), Y(1) \} \indep T ~|~ \bm{X} } E\left[Y(1)|T=1, \bm{X}\right]  - E\left[Y(0)|T = 0,\bm{X}\right] \\
        &= E\left[Y|T=1, \bm{X}\right]  - E\left[Y|T = 0,\bm{X}\right]. \notag
        \end{split}
       \label{eq:tauX}
       \]

~

\paragraph{Assumption 2 (Overlap):} \label{para:ass2}~
\\
       \[
       0 < P(T = 1 | \bm{X}) < 1, ~ for ~all ~\bm{X}.
       \]

To ensure that there are useful observations for estimating the causal effect, we assume the probability that a subject is assigned to the treatment group is limited to range between 0 and 1. \textbf{Assumption 2} also helps to ensure that the bias-correction information is available in the entire domain of $ \bm{X} $ \citep{zhao2016entropy} and equivalently requires that the covariate distribution provides sufficient overlap between the treatment and the control group.

Reminding that under Assumption 1, the ATE can be expressed as
\begin{equation}
\begin{aligned}
\tau &= E\left[ Y_i(1) - Y_i(0) \right] \\
&=E\{E\left[ Y(1) - Y(0) | \bm{X} \right]\} \\
&=E\left[ \mu_1(x)  - \mu_0(x) \right] \\
&=\int \mu_1(x) dF(x) - \int \mu_0(x) dF(x)
\end{aligned}
\label{eq:ATE}
\end{equation}
where $ F(x) $ indicates the true distribution of covariates. Within a weighting framework, we denote the weighted sample average treatment effect (SATE) with weights $\bm{w} = \{w_1, \cdots, w_N\} = \{p_1, \cdots, p_{n_1}, q_1, \cdots, q_{n_0}\}$, $ n_1 + n_0 = N $, as
$$ \hat{\tau}_N^w = \sum_{i=1}^{n}T_iw_iY_i-\sum_{i=1}^{n}(1-T_i)w_iY_i = \sum_{i=1}^{N} p_i T_i Y_{i} - \sum_{j=1}^{N} q_j (1 - T_j) Y_{j}$$ 
and the weigthed empirical distributions of the treatment and control groups as
$$ \hat{F}_{N,1}^w (x)= \sum_{i = 1}^{n_1} p_i I(X_{1i} \leq x) \quad , \quad \hat{F}_{N,0}^w (x)= \sum_{i = 1}^{n_0} q_j I(X_{0j} \leq x),$$
where $I(\cdot)$ is an indicator function. Therefore, we have
\begin{equation}
	\begin{aligned}
		\hat{\tau}^w_N-\tau=&\sum_{i=1}^{n}T_iw_iY_i-\sum_{i=1}^{n}(1-T_i)w_iY_i-\tau\\
		=&\sum_{i=1}^{n}T_iw_i(Y_i-\mu_1(X_i))-\sum_{i=1}^{n}(1-T_i)w_i(Y_i-\mu_0(X_i))+\sum_{i=1}^{n}T_iw_i\mu_1(X_i)\\
		&-\sum_{i=1}^{n}(1-T_i)w_i\mu_0(X_i)-\tau\\
		=& \left\{\sum_{i=1}^{n}T_iw_i(\mu_1(X_i)-\mu_0(X_i))-\tau\right\}+ \left\{\sum_{i=1}^{n}T_iw_i\mu_0(X_i)-\sum_{i=1}^{n}(1-T_i)w_i\mu_0(X_i)\right\}\\
		&+ \left\{\sum_{i=1}^{n}T_iw_i(Y_i-\mu_1(X_i))-\sum_{i=1}^{n}(1-T_i)w_i(Y_i-\mu_0(X_i))\right\}.
	\end{aligned}
\label{eq:bias}
\end{equation}
According to Equation (\ref{eq:bias}), we know that the bias of weighted SATE $ \hat{\tau}_N^w $ depends entirely on the three items 
   \[
   \sum_{i=1}^{n}T_iw_i(\mu_1(X_i)-\mu_0(X_i))-\tau ,
   \label{eq:term1}
   \]
   \[
   \sum_{i=1}^{n}T_iw_i\mu_0(X_i)-\sum_{i=1}^{n}(1-T_i)w_i\mu_0(X_i) ,
   \label{eq:term2}
   \]
   \[
   \sum_{i=1}^{n}T_iw_i(Y_i-\mu_1(X_i))-\sum_{i=1}^{n}(1-T_i)w_i(Y_i-\mu_0(X_i)).
   \label{eq:term3}
   \]
Here, the second item directly shows the importance of balancing the distributions of the covariates between the treated and control groups, which is the goal of our proposed weighting framework. To further ensure $ \lim_{N \rightarrow \infty} (\hat{\tau}_N^w - \tau) = 0 $, we need the following \textbf{Assumption 3}.
\paragraph{Assumption 3 (Constant Conditional ATE):}\label{para:ass3}~
\\
\[
\mu_1(\bm{X}) - \mu_0(\bm{X}) = \tau, ~ for ~all ~\bm{X}.
\]
With \textbf{Assumption 3}, we assume a constant causal effect, so the ATE is numerically the same as the ATT.

\section{Existing Weighting Methods for Covariate Balancing}\label{sec:review}
We next give a brief overview of some existing approaches.

\subsection{IPW: Inverse Probability Weighting}
According to \citet{ma2010robust}, the IPW assigns a weight
$$\widehat{p}_i^{IPW} = \frac{\frac{T_{1i}}{Np(X_{1i})}}{\sum_{i=1}^{n_1}\frac{T_{1i}}{Np(X_{1i})}}~$$ 
to $Y_{1i}$ (the outcome in the treatment group), $i = 1,2,\cdots,n_1$, and a weight
$$\widehat{q}_j^{IPW} = \frac{\frac{1-T_{0j}}{N(1-p(X_{0j}))}}{\sum_{j=0}^{n_0}\frac{1-T_{0j}}{N\left(1-p(X_{0j})\right)}}~ $$
to $Y_{0j}$ (the outcome in the control group), $j = 1,2,\cdots,n_0$, to estimate ATE, thereby deriving a natural estimator of ATE, that is,
\begin{equation}
\begin{split}
\widehat{\tau}_{~_{IPW}} &=\frac{1}{N}\sum_{i=1}^{N}\left[ \frac{T_i Y_i}{p(X_i)} - \frac{(1-T_i)Y_i}{1-p(X_i)} \right] \\
&=\frac{1}{N}\sum_{i=1}^{N}\frac{(2T_i-1)Y_i}{1-T_i+(2T_i-1)p(X_i)}\\
&=\sum_{i=1}^{n_1}\frac{T_{1i}Y_{1i}}{Np(X_{1i})} - \sum_{j=1}^{n_0}\frac{(1-T_{0j})Y_{0j}}{N(1-p(X_{0j}))}.  \notag
\end{split}
\end{equation}

Besides, a natural estimator of \textbf{ATT} using inverse probability weighting is
\begin{equation}
\begin{split}
\widehat{\tau}_{~_{1,IPW}} &=\frac{1}{n_1} \sum_{i=1}^N\left[ T_i Y_i - (1-T_i) \frac{\hat{e}(X_i)}{1-\hat{e}(X_i)}Y_i \right] \\ &=\frac{1}{n_1} \sum_{i=1}^N \left(T_i \cdot  Y_i\right) -\frac{1}{n_1}\sum_{i=1}^N \left[(1-T_i) \frac{\hat{e}(X_i)}{1-\hat{e}(X_i)}Y_i\right]  \\
&=\sum_{i=1}^{n_1}\frac{Y_{1i}}{n_1} - \sum_{j=1}^{n_0}\frac{\frac{\hat{e}(X_i)}{1-\hat{e}(X_i)}Y_{0j}}{n_1} .  \notag
\end{split}
\end{equation}
Therefore, the weights assigned to $Y_{1i},~i = 1,2,\cdots,n_1$  for estimating \textbf{ATT} is
$$\widehat{p}_{i,trt}^{IPW} = \frac{1}{n_1}~,$$ and the weights assigned to $Y_{0j},~j = 1,2,\cdots,n_0$ is
$$\widehat{q}_{j,trt}^{IPW} = \frac{\hat{e}(X_i)}{n_1\cdot \left(1-\hat{e}(X_i)\right)} ~. $$	

\subsection{EB: Entropy Balancing \citep{hainmueller2012entropy}}
The entropy balancing (EB) scheme by \citet{hainmueller2012entropy} considered:
\begin{align}
	\bm{\min \limits_{w_i}} H(w) = \sum_{i|D = 0} h(w_i) \notag
\end{align}
$$s.t. \left\{
\begin{array}{rcl}
	& \sum_{i|D = 0} w_i c_{ri}(X_i) = m_r,   & with~ r \in 1,2,\cdots,R\\
	& \sum_{i|D = 0}w_i = 1,   & w_i \geq 0 ~ for ~ all ~ i ~ such ~ that ~ D = 0\\
\end{array} 
\right.
$$
The weights $\widehat{p}_i^{EB} = \frac{1}{n_1}~ (1,\cdots,n_1)$ and $\widehat{q}_j^{EB} = w_j ~(j=1,\cdots,n_0)$ of EB method can be directly calculated with the R package ``\textbf{ebal}''.



\subsection{CBPS: Covariate Balance Propensity Score \citep{imai2014covariate}}
\citet{imai2014covariate} proposed a relatively more stable method, namely the Covariate Balance Propensity Score (CBPS), for estimating propensity score so that covariate balance is optimized. CBPS can significantly improve the poor empirical performance of propensity score matching and weighting methods. The key idea is using propensity score weighting to characterize the covariate balance:
   \[
   \mathbb{E}\left(
   \frac{T_i \tilde{\bm{X}}_i}{\pi_{\beta}(\bm{X}_i)} - \frac{(1 - T_i) \tilde{\bm{X}}_i}{1 - \pi_{\beta}(\bm{X}_i)} 
   \right) = 0
   \label{eq:CBPS}
   \]
where $ \tilde{\bm{X}}_i = f(\bm{X}_i) $ is an D-dimensional vector-valued measurable function of $ \bm{X}_i $ specified by the researcher. Equation (\ref{eq:CBPS}) holds for any function of covariates. Selecting a specific $ f(\cdot) $, for example, setting $ \tilde{X}_i =(X_i^T X_i^{2T})^{T} $, Equation (\ref{eq:CBPS}) helps to balance both the first and second moments of all covariates.

The CBPS method can be implemented through the open-source R package ``CBPS''.

\subsection{CAL: Calibration Estimator \citep{chan2016globally}}
A general class of calibration (\textbf{CAL}) estimators of ATE is established with a wide class, such as exponential tilting, of calibration weights. The weights are constructed to achieve an exact moment balance of observed covariates among the treated, the control, and the combined group. Global semiparametric efficiency has been established for the CAL estimators. 

For a fixed $g \in \mathbb{R}$, let $ D(f, g) $ denote a continuously differentiable distance measure for $ f \in \mathbb{R}$. Being nonnegative and strictly convex in $ f $, $ D(f, g) = 0 $ if and only if $ f=g $. The general idea of calibration as in \citet{deville1992calibration} is to minimize the aggregate distance between the final weights $ w = (w_1, \cdots, w_N) $ to a given vector of design weights $ d = (d_1, \cdots, d_N) $ subject to moment constraints. 
Avoiding estimating the design weights, CAL first used a vector of misspecified uniform design weights $ d^* = (1, \cdots, 1) $, and constructed the final weights $ w $ by solving the following constrained optimization problem:\begin{gather}
Minimize ~\sum_{i=1}^{N}D(w,1), \notag \\
s.t. \begin{cases}
\frac{1}{N}\sum_{i=1}^{N} T_i w_i u(X_i) = \frac{1}{N}\sum_{i=1}^{N} u(X_i) \\
\frac{1}{N}\sum_{i=1}^{N} (1-T_i) w_i u(X_i) = \frac{1}{N}\sum_{i=1}^{N} u(X_i)
\end{cases} .
\end{gather}

This CAL method can be implemented through an open-source R package ``ATE'' for estimating both the ATE and ATT.

\section{The Weighted Kernel Covariate Balancing}\label{sec:KDbal}

\subsection{Kernel Distance}\label{subsec:kd}
Let $ X_1,\cdots,X_n $ and $ Y_1,\cdots,Y_m $ be random samples from two unknown probability measures, $ \mathbb{P} $ and $ \mathbb{Q} $, then researchers are often interested in detecting the difference and estimating the discrepency between them.
In this paper, we use a probability metric with $\zeta$-structure \citep{zolotarev1983probability} 
    \[ 
    \gamma_{\mathscr{F}} (P,Q) = \sup\limits_{f \in \mathscr{F}} | \int f dP - \int f dQ |,
    \]
where $\mathscr{F}$ indicates a class of functions, to measure the distance between the treatment and the control covariates. 

As for the implementation of $ \gamma(P, Q) $, we can define $|| \cdot ||_{\mathscr{K}}$ to be a norm of a Reproducing Kernel Hilbert Spaces (RKHS) $ \mathscr{K} $ and set $\mathscr{F} = \{ f:||f||_{\mathscr{K}} \neq 1 \}$, by which we can get a ``kernelized'' version of the total variation distance. Restricting $ k $ to be a strictly positive definite kernel function corresponding to RKHS $\mathscr{K}$, then we can get a closed-form solution for $ \gamma(P,Q) $  \citep{sriperumbudur2012empirical}, that is  
     \[
     \gamma_k(P_{n_1},Q_{n_0}) = || \sum_{i=1}^N w_i k(\cdot, \bm{X}_i) ||_{\mathscr{K}} = \sqrt{\sum_{i,j=1}^N w_i w_j k(\bm{X}_i, \bm{X}_j)},
     \]
where $ P_{n_1} $ and $ Q_{n_0} $ respectively denote the empirical measure of $ \bm{X}|T=1 $ and $ \bm{X}|T=0 $, and $w_i$ is the sample weight for the unit $i \in \{1,\cdots,N\}$. Let us define $ w_i = 1/n_1 $ if $ T_i=1 $ and $ w_i = 1/n_0 $ if $ T_i=0 $, then we can get a simple as well as closed-form analytical solution for the empirical estimate of the probability metric under the assumption that the function class represents an RKHS. The computation of $ \gamma_k(P_{n_1},Q_{n_0}) $ is $ O(N^2) $ but it is also independent of the dimension of the confounders.

We can use $ \gamma_k(P_{n_1},Q_{n_0}) $ as a diagnostic for covariate balance, the smaller values of which denote better performance. 


\subsection{The Integral Probability Metric (IPM)}
Given two probability measures, $ \mathbb{P} $ and $ \mathbb{Q} $ defined on a measurable space \textit{S}, the integral probability metric (IPM) \citep{sriperumbudur2012empirical}, also called probability metrics with a $ \zeta $-structure, between $ \mathbb{P} $ and $ \mathbb{Q} $ is defined as
\begin{equation}
\gamma_{\mathcal{F}}(\mathbb{P}, \mathbb{Q}) = \sup \left\{ \left\lvert \int_\textit{S} f d\mathbb{P}  - \int_\textit{S} f d\mathbb{Q} \right\rvert: f \in \mathcal{F} \right\}
\label{eq:IPM}
\end{equation}
where $ \mathcal{F} $ is a class of real-valued bounded measurable functions on \textit{S}. 
The choice of functions is the crucial distinction between different IPMs and various popular distance measures, for example, the \textit{Kantorovich metric} and the \textit{total variation distance}, can be obtained by choosing appropriate $ \mathcal{F}$ in (\ref{eq:IPM}). When $ \mathcal{F} = \{f: || f ||_\mathcal{H} \leq 1 \} $, the corresponding $ \gamma_{\mathcal{F}} $ is called kernel distance or maximum mean discrepancy, where $ \mathcal{H} $ represents a reproducing kernel Hilbert space (RKHS) with $ k $ as its reproducing kernel (r.k.). In this situation, we can rewrite the function space $ \mathcal{F} $ as $ (\mathcal{H}, k) $ and the IPM $ \gamma_{\mathcal{F}} $ as $ \gamma_{\mathcal{F}_k} $. The kernel distance has been widely used in statistical applications, including homogeneity testing, independence testing, conditional independence testing, and mixture density estimation.

\subsection{Estimators of IPM}
Let $ \left\{X_{P,i}: ~i=1,2,\cdots,n\right\} $ and $ \left\{X_{Q,j}:~ j=1,2,\cdots,m\right\} $ be i.i.d. samples randomly drawn from $ \mathbb{P} $ and $ \mathbb{Q} $, respectively, the empirical estimator of $ \gamma_{\mathcal{F}}(\mathbb{P}, \mathbb{Q}) $ is given by
    \begin{equation}
    \gamma_{\mathcal{F}}(\mathbb{P}_{n}, \mathbb{Q}_{m}) = \sup_{f \in \mathcal{F}} \left\lvert \sum_{i=1}^N w^{E}_{i} f(X_i) \right\rvert,
    \label{eq:empiricalIPM}
    \end{equation}
where $ \mathbb{P}_{n}(x) := \sum_{i = 1}^{n} \frac{1}{n} I(X_{P,i} \leq x) $ and $ \mathbb{Q}_{m}(x) := \sum_{j = 1}^{m} \frac{1}{m} I(X_{Q,j} \leq x) $ represent the empirical distributions of $ \mathbb{P} $ and $ \mathbb{Q} $, respectively, $ I(\cdot) $ is an indicator function,  $ N = n + m$, $ w_{i}^{E} = \frac{1}{n} $ when $ X_i = X_{P,i}  $ for $ i = 1, \cdots , n $ and $ w_{j}^{E} = -\frac{1}{m} $ when $ X_j = X_{Q,j}  $ for $ j = 1, \cdots , m $.

Correspondingly, we define the weighted estimator of of $ \gamma_{\mathcal{F}}(\mathbb{P}, \mathbb{Q}) $ as
     \begin{equation}
     \gamma_{\mathcal{F}}(\mathbb{P}_{n}^{w}, \mathbb{Q}_{m}^{w}) = \sup_{f \in \mathcal{F}} \left\lvert \sum_{i=1}^N w_{i} f(X_i) \right\rvert,
     \label{eq:weightedIPM}
     \end{equation}
where $ \mathbb{P}^{w}_{n}(x) := \sum_{i = 1}^{n} p_i I(X_{P,i} \leq x) $ and $ \mathbb{Q}_{m}(x) := \sum_{j = 1}^{m} q_j I(X_{Q,j} \leq x) $ represent the wighted distributions of $ \mathbb{P} $ and $ \mathbb{Q} $, respectively, $ w_i = p_i $ when $ X_i = X_{P,i}  $ for $ i = 1, \cdots , n $ and $ w_j = -q_j $ when $ X_j = X_{Q,j}  $ for $ j = 1, \cdots , m $. $ \{p_i: i = 1, \cdots , n\} $ and $ \{q_j: j = 1, \cdots , m\} $ are the balancing weights derived frome some weighting methods.

\subsection{Weighted Estimator of the Kernel Distance}
\begin{thm}
	Let $ k $ be a strictly positive definite kernel, i.e., for all $ N \in \mathbb{N} $, $ \{ \alpha_i \}_{i = 1}^N \subset \mathbb{R} \backslash \{0\} $ and all mutually distinct $ \{ \theta_i \}_{i = 1}^N \subset S $, $ \sum_{i,j = 1}^N \alpha_i \alpha_j k(\theta_i,\theta_j) > 0$. Then, for $ \mathcal{F} = \mathcal{F}_k := \{f: \lVert f \rVert_\mathcal{H} \leq 1\}$, the following function $ f $ is the unique solution to (\ref{eq:weightedIPM}):
	\begin{equation}
	f = \frac{1}{\left\lVert \sum_{i = 1}^{N} w_i k(\cdot, X_i) \right\rVert_\mathcal{H}}  \sum_{i = 1}^{N} w_i k(\cdot, X_i)
	\label{eq:f}
	\end{equation}
	and the corresponding weighted estimator of the kernel distance is
	\begin{equation}
	\gamma_{k,N}(\mathbb{P}_{n},\mathbb{Q}_{m}) = \left\lVert \sum_{i = 1}^N w_i k(\cdot, X_i) \right\rVert = \sqrt{\sum_{i,j = 1}^N w_i w_j k(X_i,X_j) }.
	\label{eq:wkd}
	\end{equation}
\end{thm}

\noindent \textit{\textbf{Proof.}} Consider $ \gamma_{k,N}(\mathbb{P}_{n},\mathbb{Q}_{m}):=\sup\left\{ \sum_{i = 1}^N w_i f(X_i) : \lVert f \rvert_\mathcal{H} \leq 1 \right\} $, which can be written as
\[
\gamma_{k,N}(\mathbb{P}_{n},\mathbb{Q}_{m}) = \sup \left\{ \left \langle  f,\sum_{i = 1}^{N} w_i k(\cdot, X_i) \right \rangle  _\mathcal{H} : \lVert f \rVert_\mathcal{H} \leq 1 \right\},
\]
where  we have used the reproducing property of $ \mathcal{H} $, i.e., $\forall f \in \mathcal{H}, \forall x \in \textit{S}, f(x) = \langle f,k(\cdot, x)\rangle_\mathcal{H}  $.

Following the Cauchy-Schwartz inequality, which states that for all vectors $ u $ and $ v $ of an inner product space it is true that
\begin{equation}
	\lVert \langle u,v \rangle \rVert \leq \lVert u \rVert \cdot \lVert v \rVert
	\label{eq:csi}
\end{equation}
where $ \langle \cdot , \cdot \rangle $ is the inner product, we have
\[
\begin{aligned}
	\gamma_{k,N}(\mathbb{P}_{n},\mathbb{Q}_{m}) :=\sup\left\{ \sum_{i = 1}^N w_i f(X_i) : \lVert f \rvert_\mathcal{H} \leq 1 \right\} &\leq \sup\left\{ \left\lVert f \right\rVert_\mathcal{H} \cdot \left\lVert \sum_{i = 1}^{N} w_i k(\cdot, X_i) \right\rVert_\mathcal{H}: \lVert f \rVert_\mathcal{H} \leq 1 \right\} \\
	&\leq \sup\left\{ \left\lVert \sum_{i = 1}^{N} w_i k(\cdot, X_i) \right\rVert_\mathcal{H} \right\}
\end{aligned}
\]
Therefore, we can set 
\[
f = \frac{1}{\left\lVert \sum_{i = 1}^{N} w_i k(\cdot, X_i) \right\rVert_\mathcal{H}}  \sum_{i = 1}^{N} w_i k(\cdot, X_i) .
\]

Since $ k $ is strictly positive definite, $ \gamma_{k,N}(\mathbb{P}_{n},\mathbb{Q}_{m}) = 0 $  if and only if $ \mathbb{P}_{n} = \mathbb{Q}_{m} $, which therefore ensures that (\ref{eq:f}) is well-defined.

\subsection{The Gaussian Kernel and the Information Matrix} \label{subsec:Gkd}
According to \citet{rasmussen2003gaussian}, one of the choice of kernel $ k $ for covariate balance is the Gaussian kernel, that is
       \[
       K(\bm{x},\bm{x'};\sigma^2) = exp\left( \frac{-|| \bm{x} -\bm{x'} ||^2}{\sigma^2} \right),
       \]
where the parameter $ \sigma $ determines the width of the Gaussian kernel and it can either be fixed or estimated from the data. In this article, we estimate $ \sigma $ with the median of all possible pairwise squared Euclidean distances between all pairs of subjects. 
       \[
       \sigma = median \begin{Bmatrix}
       \{
       	(\bm{X}_{1i} - \bm{X}_{1j})^2_{i,j \in trt}, ~
       	(\bm{X}_{1i} - \bm{X}_{0j})^2_{i \in trt, j~ \in col},
       	(\bm{X}_{0i} - \bm{X}_{0j})^2_{i,j \in col}
       \}
       ~ > ~0 
       \end{Bmatrix}
       \]
The Information Matrix $ \bm{K}_G $ corresponding to the Gaussian kernel $   K(\bm{x},\bm{x'};\sigma^2) $ is
       \begin{equation}
          \begin{aligned}
          \bm{K}_G & = \begin{pmatrix}
          K_1 & -K_{10} \\
          -K_{01} & K_0 \\
          \end{pmatrix} \\
          & = \begin{pmatrix}
          \left(K(X_{1i},X_{1j})\right)_{n_1 \times n_1} & -\left(K(X_{1i},X_{0j})\right)_{n_1 \times n_0} \\
          \left(-\left(K(X_{1i},X_{0j})\right)_{n_1 \times n_0}\right) ^T& 
          \left( -K(X_{0i},X_{0j})\right)_{n_0 \times n_0}
          \end{pmatrix} .
          \end{aligned}
       \label{eq:Kmat}
       \end{equation}


\subsection{The Kernel-distance-based (KDB) Covariate Balance} \label{subsec:wkdbal}
In this article, we propose a preprocessing procedure, the kernel-distance-based covariate balancing method, to create balanced samples for evaluating the treatment effects. In this procedure, we create two groups of scalar weights for the treatment and the control groups so that the reweighted treatment and control groups can match precisely at the specified moments. 

\subsubsection{ATE Estimation}\label{subsubsec:KDATE}
In this section, we introduce the estimation of ATE
  \[
  \tau \equiv E[Y(1)- Y(0)]
  \]
with the weighted mean difference between the treatment and the control outcomes, that is
  \[
  \hat{\tau} = \sum_{i=1}^{n_1} p_i \cdot Y_{1i} - \sum_{j=1}^{n_0} q_j \cdot Y_{0j}.
  \]
The weights $ \{p_i:i = 1,\cdots, n_1 \} $ and $ \{q_j:j = 1,\cdots, n_0 \} $ derive from the following KDB reweighting scheme:
\begin{equation}
    \label{eq:ATEKDopt}
	\begin{aligned}
	\min \limits_{\bm{\vec{p}},\bm{\vec{q}}} r^W(\bm{\vec{p}},\bm{\vec{q}})  = & \sum_{i=1}^{N}\sum_{j=1}^{N}p_i p_j K({X}_{1i},{X}_{1j}) \cdot I(i,j \in \textbf{trt})  +  \\
	& \sum_{i=1}^{N}\sum_{j=1}^{N}q_i q_j K(\bm{X}_{0i},\bm{X}_{0j}) \cdot I(i,j \in \textbf{col}) - \\
	&2 \sum_{i=1}^{N}\sum_{j=1}^{N}p_i q_j K(\bm{X}_{1i},\bm{X}_{0j}) \cdot I(i \in \textbf{trt}, j \in \textbf{col})
	\end{aligned}
\end{equation}
  \[
  s.t. \left\{
  \begin{array}{rcl}
  & \sum_{i=1}^{n_1}p_i = 1,   & p_i \geq 0\\
  & \sum_{j=1}^{n_0}q_j = 1,   & q_j \geq 0\\
  & \sum_{i=1}^{n_1}p_i\bm{g}(\bm{X}_{1}) =  \sum_{j=1}^{n_0}q_j\bm{g}(\bm{X}_{0})  &\\ 
  \end{array} 
  \right.
  \]
where $I(\cdot)$ is an indicator function, and $\bm{g}(\cdot)$ indicates a set of constraint functions imposed on the covariate moment. Here, for convenience, we set $\bm{g}(\cdot) = \{g_{1}(\cdot), \cdots, g_{D}(\cdot)\}$ to be the sample mean function at all covariates so that the reweighted treatment and control groups can match on the first moment, that is 
     \[
     g_{d}(\bm{X}_{t}) := \frac{\sum_{i=1}^{n_t} X_{ti,d}}{n_{t}},
     \]
where $ t \in \{0,1\}, i \in \{1, \cdots, n_{t}\}, d \in \{1,\cdots, D\} $. 
To express concisely, we define $ \vec{\bm{w}} = (p_{1},\cdots,p_{n_1},q_{1},\cdots,q_{n_0})^T $, $ \bm{\vec{b_0}}_{_{(2+D+N) \times 1}} = \begin{pmatrix} 1 & 1 & 0 & 0 & \cdots &0 \end{pmatrix}^T $, and 
\begin{align}
\bm{A}^T & =\begin{pmatrix}
\vec{\bm{1}}_{1\times n_1},&~ &\vec{\bm{0}}_{1\times n_0}  \\
\vec{\bm{0}}_{1\times n_1},&~ &\vec{\bm{1}}_{1\times n_0}  \\
\vec{\bm{X}}_{1,trt},&~ &-\vec{\bm{X}}_{1,col}  \\
\vdots,&~ &\vdots  \\
\vec{\bm{X}}_{D,trt},&~ &-\vec{\bm{X}}_{D,col}  \\
&~ \bm{I}_{N \times N} &~
\end{pmatrix}_{(2+D+N) \times N}  = \begin{pmatrix}
1_{1} & \cdots & 1_{n_1} & 0_1 & \cdots & 0_{n_0} \\
0_{1} & \cdots & 0_{n_1} & 1_1 & \cdots & 1_{n_0} \\
X_{1,11} & \cdots & X_{1,1n_1} & -X_{1,01} & \cdots & -X_{1,0n_0} \\
\vdots &\vdots &\vdots & \vdots & \vdots & \vdots\\
X_{D,11} & \cdots & X_{D,1n_1} & -X_{D,01} & \cdots & -X_{D,0n_0} \\
1 & 0 & 0 & \cdots & 0 & 0 \\
0 & 1 & 0 & \cdots & 0 & 0  \\
\vdots &\vdots &\vdots & \ddots & \vdots & \vdots\\
0 & 0 & 0 & \cdots & 1 & 0 \\
0 & 0 & 0 & \cdots & 0 & 1 \\
\end{pmatrix}_{(2+D+N) \times N} , \notag
\end{align}
where $D$ refers to the total number of covariates, $ \vec{\bm{X}}_{d,trt} $ and $ \vec{\bm{X}}_{d,col} $ respectively on behalf of the $ d^{th} $ covariate in the treatment and control group, and then the optimization problem for estimating ATE can be expressed as 
\begin{equation}
\label{eq:ATEKDoptv2}
\begin{aligned}
\min \limits_{\vec{\bm{w}}} r^W(\vec{\bm{w}})  = \vec{\bm{w}}^T \bm{K}_G \vec{\bm{w}}
\end{aligned}
\end{equation}
$$ s.t.~ \bm{A}^T \vec{\bm{w}} \geq \bm{\vec{b}_0} $$
Here, the first $2+D$ constraints are treated as equality constraints, and all further as inequality constraints.

\subsubsection{Stable Weight Estimation of ATE}\label{subsubsec:stableKDATE} 
To further construct stable weights for estimating the ATE, we can make a small modification to the optimization problem in (\ref{eq:ATEKDoptv2}). Compared to the reweighting scheme in section \ref{subsubsec:KDATE}, we only add a tuning parameter $\lambda$ to the diagonal of $ \bm{K}_G $ in order to control the variance of $ \hat{\tau} $, while keeping all the other settings and parameters the same. Defining $ \vec{\bm{w_0}} = \vec{\bm{w}}|_{p_i =\frac{1}{n_1}, q_j = \frac{1}{n_0}} =  (\frac{1}{n_1},\cdots,\frac{1}{n_1},\frac{1}{n_0},\cdots,\frac{1}{n_0})^T $,  the reweighting scheme for stable ATE estimation is:
\begin{equation}
\label{eq:sATEKDopt}
\begin{aligned}
\min \limits_{\vec{\bm{w}}} r^W(\vec{\bm{w}}) = \vec{\bm{w}}^T \bm{K}_G \vec{\bm{w}} + \lambda \left( \vec{\bm{w}} - \vec{\bm{w_0}} \right)^T\left( \vec{\bm{w}} - \vec{\bm{w_0}} \right)
\end{aligned}
\end{equation}
$$ s.t.~ \bm{A}^T \vec{\bm{w}} \geq \bm{\vec{b}_0} $$
It is easy to prove that the optimization problem (\ref{eq:sATEKDopt}) satisfying constraints above is equivalent to 
\begin{equation}
\label{eq:sATEKDoptv2}
\begin{aligned}
\min \limits_{\vec{\bm{w}}} r^W(\vec{\bm{w}}) = \vec{\bm{w}}^{T} (\bm{K}+\lambda \bm{I}_{_{N \times N}}) \vec{\bm{w}}
\end{aligned}
\end{equation}
$$ s.t.~ \bm{A}^T \vec{\bm{w}} \geq \bm{\vec{b}_0} $$

When $\lambda = 0$, this optimization problem (\ref{eq:sATEKDoptv2}) degenerates into problem (\ref{eq:ATEKDoptv2}) and it controls bias, but not variance of ATE estimation. When $\lambda$ increases, this optimization problem may increase bias while reduce the variance of ATE estimation.

\subsubsection{ATT Estimation}\label{subsubsec:KDATT} 
In this section, we introduce the KDB reweighting scheme for estimating the ATT
   \[
   \tau_1 := E[Y(1)- Y(0) | T = 1] = \mu_{1|1} - \mu_{0|1}.
   \]
We estimate $ \tau_1 $ with the mean difference between the treatment outcomes and the weighted control outcomes, that is, 
\[
\hat{\tau}_1 = \sum_{i=1}^{n_1} \frac{1}{n_1} \cdot Y_{1i} - \sum_{j=1}^{n_0} q_j \cdot Y_{0j},
\]
where $ \{q_j:j = 1,\cdots, n_0 \} $ derive` from the following reweighted scheme:
\begin{equation}
\label{eq:ATTKDopt}
\begin{aligned}
\min \limits_{\bm{\vec{q}}} r^W(\bm{\vec{q}})  = & \sum_{i=1}^{N}\sum_{j=1}^{N}\frac{1}{n_1} \frac{1}{n_1} K(\bm{X}_{1i},\bm{X}_{1j}) \cdot I(i,j \in \textbf{trt})  +  \\
& \sum_{i=1}^{N}\sum_{j=1}^{N}q_i q_j K(\bm{X}_{0i},\bm{X}_{0j}) \cdot I(i,j \in \textbf{col}) - \\
&2 \sum_{i=1}^{N}\sum_{j=1}^{N} \frac{1}{n_1} q_j K(\bm{X}_{1i},\bm{X}_{0j}) \cdot I(i \in \textbf{trt}, j \in \textbf{col})
\end{aligned}
\end{equation}
\[
s.t. \left\{
\begin{array}{rcl}
& \sum_{j=1}^{n_0}q_j = 1,   & q_j \geq 0\\
& \sum_{i=1}^{n_1}\frac{1}{n_1}\bm{g}(\bm{X}_{1}) =  \sum_{j=1}^{n_0}q_j\bm{g}(\bm{X}_{0})  &\\ 
\end{array} 
\right.
\]
Just like the ATE estimation, we define $ \bm{\vec{q}}_{_{n_0 \times 1}} = \left(q_1,~ q_2,~ \cdots,~ q_{n_0} \right)^T, $
and reset $ \vec{\bm{w}}$, $ \bm{\vec{b_0}}_{_{(1+D+n_0)  \times 1}} $, $ \bm{A}^T $ respectively to be 
   \[
    \vec{\bm{w}} = (p_{1},\cdots,p_{n_1},q_{1},\cdots,q_{n_0})^T ,
   \]
   
   \[ 
   \bm{\vec{b_0}}_{_{(1+D+n_0)  \times 1}} = \begin{pmatrix}
   1  & \overline{X}_{1,1} & \overline{X}_{2,1} &\cdots &\overline{X}_{D,1} & 0 & \cdots &0
   \end{pmatrix}^T
   \]

\begin{align}
\bm{A}^T & =\begin{pmatrix}
\vec{\bm{1}}_{1\times n_0}  \\
\bm{X}_{1,0} \\
\vdots \\
\bm{X}_{D,0} \\
\bm{I}_{n_0 \times n_0}
\end{pmatrix}_{(1+D+n_0) \times n_0} 
= \begin{pmatrix}
1_1 & 1_2 & \cdots & 1_{n_0} \\
X_{1,01} &X_{1,02} & \cdots & X_{1,0n_0} \\
\vdots & \vdots & \ddots & \vdots\\
X_{D,01} &X_{D,02} & \cdots & X_{D,0n_0} \\
1 & 0 &\cdots & 0 \\
0 & 1 &\cdots & 0 \\
\vdots & \vdots & \ddots & \vdots\\
0 & 0 & \cdots & 1 \\
\end{pmatrix}_{(1+D+n_0) \times n_0}  , \notag
\end{align}
where $ p_i = \frac{1}{n_1},~ i \in \{1, \cdots, n_1\} $ and $ \overline{X}_{d,1},~d \in \{1,\cdots,D\}  $ refers to the sample mean of the $ d^{th} $ covariate in the treatment group. Therefore, we can also simplify the ATT optimization problem above into a matrix form
\begin{equation}
\label{eq:ATTKDoptv2}
\begin{aligned}
\min \limits_{\vec{\bm{w}}} r^W  (\vec{\bm{w}})  &= \vec{\bm{w}}^{~T} \bm{K}_G~ \vec{\bm{w}} \\
s.t.~ \bm{A}^T \vec{\bm{q}} &\geq \bm{\vec{b}_0}
\end{aligned}
\end{equation} 
Here, the first $1+D$ constraints are treated as equality constraints, and all further as inequality constraints.

\subsection{Propositions of the KDB Covariate Balance}\label{subsec:props4kd}
\begin{pro}~
	\\
	The Gaussian Information Matrix $ \bm{K}_G $ is positive semi-definite.
\end{pro}

\noindent \textit{\textbf{Proof.}} Based on the propertied of kernel, i.e, $K(x_1,x_2)=<\varphi(x_1),\varphi(x_2)>$, for any $v\in\mathbb{R}^{N}$, we have
\begin{align*}
	v'\mathbf{K}v=&\sum_{i,j=1}^{N}v_iv_jK_{ij}\\
	=&\sum_{i,j=1}^{n_1}v_{i}v_{j}<\varphi(X_{i}),\varphi(X_{j})>+\sum_{i,j=n_1}^{N}v_{i}v_{j}<\varphi(X_{i}),\varphi(X_{j})>-2\sum_{i=1}^{n_1}\sum_{j=n_1}^{N}v_{i}v_{j}<\varphi(X_{i}),\varphi(X_{j})>\\
	=&<\sum_{i=1}^{n_1}\varphi(X_i),\sum_{j=1}^{n_1}\varphi(X_j)>+<\sum_{i=n_1}^{N}\varphi(X_i),\sum_{j=n_1}^{N}\varphi(X_j)>-2<\sum_{i=1}^{n_1}\varphi(X_i),\sum_{j=n_1}^{N}\varphi(X_j)>\\
	=&||\sum_{i=1}^{n_1}v_i\varphi(X_i)-\sum_{j=n_1}^{N}v_j\varphi(X_j)||^2\geq 0  
\end{align*}
\hfill \qedsymbol\\
as required.

\noindent \textbf{Optimization Problem:}\\
We aim to solve the following minimization problem:
\begin{equation}
\begin{aligned}
\min\limits_{x\in\mathbb{R}^N} f_0(x)=\frac{1}{2}x'\mathbf{K}x,\\
\text{st.} \mathbf{A}x=b,\\
\mathbf{B}x\leq \mathbf{0}_N,
\label{eq:op} 
\end{aligned}
\end{equation}
where $\mathbf{A}=\left( \begin{array}{cc} \mathbf{1}^\tau_{n_1} & \mathbf{0}^\tau_{n_0} \\
\mathbf{0}^\tau_{n_1} & \mathbf{1}^\tau_{n_0} \end{array} \right)$, $\mathbf{B}=-\mathbf{I}$ and $b=(1,1)^\tau$.

~

\begin{pro}
	(\ref{eq:op}) is a standard convex optimization problem.
\end{pro}

%

\newpage
\noindent \textbf{The Dual of Above Convex Problem:}\\
The Lagrange function is given by
$$  \mathcal{L}(x,\lambda,\nu)=f_0(x)+\lambda^\tau \mathbf{B}x + \nu^\tau(\mathbf{A}x-b) $$
and the dual function is given by
$\textsl{g}(\lambda,\nu)=\inf\limits_{x\in\mathbb{R}^N}\mathcal{L}(x,\lambda,\nu)$. Note
\begin{equation}
\begin{aligned}
\mathcal{L}(x,\lambda,\nu)=&
\frac{1}{2}x'\mathbf{K}x+\lambda^\tau \mathbf{B}x + \nu^\tau(\mathbf{A}x-b)\\
=& \frac{1}{2}x'\mathbf{K}x+(\lambda^\tau \mathbf{B} + \nu^\tau\mathbf{A})x-\nu^\tau b.
\end{aligned}
\end{equation}
The dual problem of (\ref{eq:op}) is $d^{*}=\sup\limits_{\lambda\geq \mathbf{0}_N,\nu}\textsl{g}(\lambda,\nu)$. Let $p^{*}$ be the optimal value of (\ref{eq:op}), we have $d^{*}\leq p^{*}$.
Since the inequality constraints are all affine, from the weak Slater condition we know that
$d^{*}=p^{*}$. Then from KKT conditions we have, \begin{align}
\mathbf{K}x+\mathbf{B}^\tau \lambda + \mathbf{A}^\tau \nu=0
\label{eq:kkt1}
\end{align}
and
$$ \mathbf{A}x=b. $$
Assuming $\mathbf{K}$ is positive difinite. Solving (\ref{eq:kkt1}) we have $x=-\mathbf{K}^{-1}(\mathbf{B}^\tau \lambda + \mathbf{A}^\tau \nu)$. Therefore,
$$
\textsl{g}(\lambda,\nu)=-\frac{1}{2}(\mathbf{B}^\tau \lambda + \mathbf{A}^\tau \nu)^\tau\mathbf{K}^{-1}(\mathbf{B}^\tau \lambda + \mathbf{A}^\tau \nu)-\nu^\tau b. 
$$
The dual problem is 
\begin{equation}
\begin{aligned}
\max_{\lambda,\nu}-\frac{1}{2}(\mathbf{B}^\tau \lambda + \mathbf{A}^\tau \nu)^\tau\mathbf{K}^{-1}(\mathbf{B}^\tau \lambda + \mathbf{A}^\tau \nu)-\nu^\tau b\\
\text{s.t.}\lambda_i\geq 0,\quad i=1,\cdots,N.
\label{eq:dual prob}
\end{aligned}
\end{equation}
Solving for $\nu$ we have
\begin{equation}
\begin{aligned}
0={\triangledown}_{\nu}g(\lambda,\nu)=-\mathbf{A}\mathbf{K}^{-1}\mathbf{A}^\tau\nu-\mathbf{A}\mathbf{K}^{-1}\mathbf{B}^\tau\lambda-b\quad\Rightarrow \quad \nu=-(\mathbf{A}\mathbf{K}^{-1}\mathbf{A}^\tau)^{-1}\left(\mathbf{A}\mathbf{K}^{-1}\mathbf{B}^\tau\lambda+b\right).
\label{eq:nu}
\end{aligned}
\end{equation}
Substitute $\nu$ with (\ref{eq:nu}), (\ref{eq:dual prob}) becomes
\begin{equation}
\begin{aligned}
\max_{\lambda_i\geq 0}\frac{1}{2}\lambda^\tau\left(\mathbf{B}\mathbf{K}^{-1}\mathbf{A}^\tau(\mathbf{A}\mathbf{K}^{-1}\mathbf{A}^\tau)^{-1}\mathbf{A}\mathbf{K}^{-1}\mathbf{B}^\tau-\mathbf{B}\mathbf{K}^{-1}\mathbf{B}^\tau\right)\lambda+b^\tau(\mathbf{A}\mathbf{K}^{-1}\mathbf{A}^\tau)^{-1}\mathbf{A}\mathbf{K}^{-1}\mathbf{B}^\tau\lambda.
\label{eq:dual nprob}
\end{aligned}
\end{equation}
Suppose $\mathbf{K}^{-1}=\left( \begin{array}{cc} \mathbf{M}_1 & -\mathbf{M}_{10} \\
-\mathbf{M}_{01} & \mathbf{M}_{0} \end{array} \right)$, we have $\mathbf{A}\mathbf{K}^{-1}\mathbf{A}^\tau=\mathbf{1}^\tau\mathbf{M}_1\mathbf{1}+\mathbf{1}^\tau\mathbf{M}_0\mathbf{1}$. For simplicity, substitute $B$ with $-\mathbf{I}$ and write $\mathbf{Q}=\mathbf{1}^\tau\mathbf{M}_1\mathbf{1}+\mathbf{1}^\tau\mathbf{M}_0\mathbf{1}$, (\ref{eq:dual nprob}) is converted to
\begin{equation}
\begin{aligned}
\max_{\lambda_i\geq 0}\frac{1}{2}\lambda^\tau\left(\mathbf{K}^{-1}\mathbf{A}^\tau\mathbf{Q}^{-1}\mathbf{A}\mathbf{K}^{-1}-\mathbf{K}^{-1}\right)\lambda+b^\tau\mathbf{Q}^{-1}\mathbf{A}\mathbf{K}^{-1}\lambda.
\label{eq:dual nnprob}
\end{aligned}
\end{equation}

\section{Simulation Studies}\label{sec:simNrda}
In this section, we perform simulation studies for evaluating the performance of different weighting methods, including the inverse probability weighting (IPW), the covariate balancing propensity score (CBPS) \citep{imai2014covariate}, the entropy balance (EB) \citep{hainmueller2012entropy}, the calibration estimator (CAL) \citep{chan2016globally} and our proposed KDB Covariate Balance. 
We use the oracle and unadjusted (UnAD) estimators as benchmarks.
In all the following simulation experiments, we assume a constant causal effect, so the ATE is numerically the same as the ATT.
For each simulation experiment, we construct $N_{sim}$ Monte Carlo datasets.   

In addition to simulation studies, we also conduct a real data analysis with the `National supported work' (NSW) dataset.

\subsection{Evaluation Metrics and Balancing Statistics} \label{subsec:embs}
To evaluate the performance of estimating ATE or ATT, we use the metric \textbf{Bias}, \textbf{\% Bias}, empirical standard deviation \textbf{(SD)}, and \textbf{RMSE}. 

\textbf{Bias} is the difference between the average estimate and the true value of ATE/ATT. \textbf{\% Bias} is the bias as a percentage of the estimate's standard deviation. A useful rule of thumb \citep{kang2007demystifying} is that the performance of interval estimates and test statistics begins to deteriorate when the bias of the point estimate exceeds about 40\% of its standard deviation. 
The empirical \textbf{SD} for an estimator is the square root of average squared differences between all estimates from the mean. Here, we use the empirical \textbf{SD} corresponding to the average value of ATE/ATT estimates, and thus it is the empirical standard deviation of all estimated values divided by the square root of sample size $ N_{sim} $. 
\textbf{RMSE} is the root-mean-square error for measuring the differences between values predicted by a weighting method and the observed values. 

To further compare the power of balancing covariates, we use the balancing statistics \bm{$r^W$}, kernel distance \textbf{(KD)}, the maximum, average, and median value of absolute standardized mean difference \textbf{(maxASMD, meanASMD, medASMD)}, the average Kolmogorov-Smirnov statistic  \textbf{(meanKS)} and the average T statistic  \textbf{(meanT)}. We introduce each balancing statistic with more details as follows. 

For every set of sample weights $\{\widehat{p}_i\}(i=1,2,\cdots,n_1)$ and $\{\widehat{q}_j\}(j=1,2,\cdots,n_0)$ obtained by different balancing methods for each Monte Carlo dataset, we calculate \bm{$r^W$} as
\begin{equation}
    \label{eq:rw}
	\begin{aligned}
	\bm{\widehat{r}^W(\vec{p},\vec{q})}  = & \sum_{i=1}^{N}\sum_{j=1}^{N}\widehat{p}_i \widehat{p}_j K(X_{1i},X_{1j}) \cdot I(i,j \in \textbf{trt})   + \\
	&\sum_{i=1}^{N}\sum_{j=1}^{N}\widehat{q}_i \widehat{q}_j K(X_{0i},X_{0j}) \cdot I(i,j \in \textbf{col}) - \\
	&2\sum_{i=1}^{N}\sum_{j=1}^{N}\widehat{p}_i \widehat{q}_j K(X_{1i},X_{0j}) \cdot I(i \in \textbf{trt}, j \in \textbf{col}).
	\end{aligned}
\end{equation}
For a simulation experiment with a total of $N_{sim}$ iterations, we calculate the average value of all $\bm{\widehat{r}^W(\vec{p},\vec{q})}$ statistics of each balancing method and compare them. \textbf{KD}, the square root of \bm{$r^W$}, has been widely used in statistical applications \citep{sriperumbudur2012empirical} and can be regarded as a reasonable measure of covariates balance \citep{xie2018causal}.

In estimating ATE, the \textbf{ASMD} of covariate $X_d ~(d = 1,\cdots,D)$ with balancing weights $\widehat{p}_i ~(i = 1,\cdots,n_1)$ and $\widehat{q}_j ~(j = 1,\cdots,n_0)$ is defined as \citep{zhang2019balance}:
   \[
   \bm{ASMD}_{d,ATE} = \frac{|\bar{X}_{d,1}^p - \bar{X}_{d,0}^q |}{\sqrt{(S_{d,1}^2+S_{d,2}^2)/2}},
   \]
where $\bar{X}_{d,1}^p$ and $\bar{X}_{d,0}^q$ respectively refer to the weighted sample mean of covariate $X_d$ in the treatment and control group
$$ \bar{X}_{d,1}^p = \sum_{i=1}^{n_1} \widehat{p}_i X_{i,d,1}, ~ \bar{X}_{d,0}^q = \sum_{j=1}^{n_0} \widehat{q}_j X_{j,d,0},$$
and $ S_{d,1}^2 $ and $ S_{d,0}^2 $ denote the sample variance of covariate $X_k$ in the treatment and control group without any adjustment
$$S_{d,1}^2 = \frac{\sum_{i=1}^{n_1}X_{i,d,1} - \bar{X}_{d,1}}{n_1-1},~ S_{d,0}^2 = \frac{\sum_{j=1}^{n_0}X_{i,d,0} - \bar{X}_{d,0}}{n_0-1}$$
In estimating ATT, the \textbf{ASMD} of covariate $X_d ~(d = 1,\cdots,D)$ with balancing weights $\widehat{p}_i = \frac{1}{n_1} ~(i = 1,\cdots,n_1)$ and $\widehat{q}_j ~(j = 1,\cdots,n_0)$ is defined as \citep{xie2019model}:
    \[
    \bm{ASMD}_{d,ATT} = | \frac{\bar{X}_{d,1}^p}{sd(X_{d,1})} - \frac{\bar{X}_{d,0}^q}{sd(X_{d,1})} |,
    \]
where $ sd(X_{d,1}) $ denotes the sample standard deviation of covariate $X_d$ in the treatment group without any adjustment
   \[
   sd(X_{d,1}) = \sqrt{\frac{\sum_{i=1}^{n_1}X_{i,d,1} - \bar{X}_{d,1}}{n_1 - 1}}.
   \]
For each Monte Carlo dataset, we calculte the \textbf{ASMD} values over all $D$ covariates and mark down the maximum, mean and median values. 

A Kolmogorov-Smirnov test can be applied to test whether two underlying one-dimensional probability distributions significantly differ and the \textbf{KS} statistic quantifies a distance between the empirical distribution functions of two samples. We calculate the \textbf{KS} statistic between the weighted samples in treatment and control group over all $D$ covariates and take the average as one balancing statistic.
$$\bm{mean KS} = \frac{\sum_{d=1}^D KS_{d}}{D}$$
where $KS_{d} =\sup\limits_x{|F_{1,n1}(X_{d,1}^p) -  F_{0,n0}(X_{d,0}^q)|}$. $ F_{1,n1}(X_{d,1}^p) $ and $ F_{0,n0}(X_{d,0}^q) $ are the empirical distribution functions of the weighted sample $\{\widehat{p}_i \cdot X_{d,1i}\}(i=1,2,\cdots,n_1)$ and $\{\widehat{q}_j\cdot X_{d,0j}\}(j=1,2,\cdots,n_0)$ respectively, and $\bm{\sup}$ is the supremum function.

A t-test is commonly applied to determine if the means of two sets of data are significantly different from each other. We conduct a Welch's t-test between the weighted means in the treatment and control group over all $D$ covariates and take the average as one balance statistic.
$$\bm{mean T} = \frac{\sum_{d=1}^D T_{d}}{D}$$
where $T_{d} = \frac{\bar{X}_{d,1}^p - \bar{X}_{d,0}^q}{S_{\bar{\triangle},d}}$, and $S_{\bar{\triangle},d} = \sqrt{\frac{S^2_{X_{d,1}^p}}{n_1}+\frac{S^2_{X_{d,0}^q}}{n_0}}$. $S^2_{X_{d,1}^p}$ and $ S^2_{X_{d,0}^q} $ are the unbiased estimators of the variances of the two weighted samples $\{\widehat{p}_i \cdot X_{d,1i}\}(i=1,2,\cdots,n_1)$ and $\{\widehat{q}_j\cdot X_{d,0j}\}(j=1,2,\cdots,n_0)$ respectively.

\subsection{Benchmark Estimators}
In different simulation scenarios, sample oracle estimators, sample unadjusted estimators, and their biases will be served as benchmarks for comparing different weighting methods.

\subsubsection{Oracle Estimator}
The sample oracle estimator of the ATE and ATT are respectively defined as 
   \[
   \widehat{\tau}_{~_{Oracle}} =\sum_i^{N} Y_{i}(1) -\sum_i^{N} Y_{i}(0),\]  
   \[\widehat{\tau}_{~_{1,Oracle}} =\sum_i^{n_1} \left[Y_{i}(1) | T_i=1\right] -\sum_i^{n_1} \left[Y_{i}(0) | T_i=1\right].
   \]
	
\subsubsection{UnAD: Unadjusted Estimator}
Without any adjustment, we assigned weight $\widehat{p}_i^{UnAD} = \frac{1}{n_1}~(i=1,2,\cdots, n_1)$ for each unit in the treatment group, and $\widehat{q}_j^{UnAD} = \frac{1}{n_0}~(j=1,2,\cdots, n_0)$ for each unit in the control group. From this, we can get the sample unadjusted estimator of the ATE, that is  $$\widehat{\tau}_{~_{UnAD}} =\sum_i^{n_1}p_i^{UnAD} Y_{1i} -\sum_j^{n_0} q_j^{UnAD} Y_{0j}.$$ 

%

\subsection{Simulation Studies}\label{subsec:sim}
For convenience, we denote
\begin{itemize}\item  KDBC: 
	
	the proposed KDB covariate balancing with only the constraints:
	\[
	\begin{aligned}
	\sum_i^{n_1}p_i = 1,\quad p_i \geq 0 ;\sum_j^{n_0}q_j = 1,\quad q_j \geq 0
	\end{aligned}
	\]
	
	\item  KDM1: 
	
	the proposed KDB covariate balancing with the constraints:
	\[
	\begin{aligned}
	&\sum_{i=1}^{n_1}p_i = 1,\quad p_i \geq 0 ;\quad \sum_{j=1}^{n_0}q_j = 1,\quad q_j \geq 0 \\
	&\sum_{i=1}^{n_1}p_i X_{k,1i} = \sum_{j=1}^{n_0}q_j X_{k,0j},\quad k \in {1,\cdots,K} \\
	\end{aligned}
	\]
\end{itemize}
\subsubsection{Simulation 1: Kang and Schafer \citep{kang2007demystifying,josey2020framework}}\label{subsubsec:sim1}
In this section, we follow the simulation example in \citet{josey2020framework}, an extensive set of experimental scenarios adapted from \citet{kang2007demystifying}, to compare different weighting methods. 

We first generate the observed covariates $\bm{X} = (X_1, X_2, X_3, X_4)^T$, where $ X_d \sim \mathcal{N}(0,1),~d = 1,\cdots,4$, and then make the following transformations to build the hidden variables $\bm{U} = (U_1,U_2,U_3,U_4)^T$
   \[
   \begin{aligned}
   U_1 &= exp(\frac{X_1}{2}), \\
   U_2 &= \frac{X_2}{1+exp(X_1)}+10, \\
   U_3 &= (\frac{X_1 \cdot X_3}{25}+0.6)^{3}, \\
   U_4 &= (X_2+X_4+20)^2.
   \end{aligned}
   \]
In subsequent applications, we standardize $\bm{U} = (U_1,U_2,U_3,U_4)^T$ so that it has a mean of zero and marginal variances of one. The propensity score model is defined as
  \[
  \pi_i^{(\delta_T)} = \frac{exp[\eta_i^{(\delta_T)}]}{1+exp[\eta_i^{(\delta_T)}]} ~ \quad \Rightarrow \quad T_i^{(\delta_T)} \sim Bin(1,\pi_i^{(\delta_T)})
  \]
where $ \delta_T\in\{X,U\} $ refers to the generative process that determines the treatment assignment. Here, $ \delta_T = X $ indicates the log odds of the propensity score is linear with $ \eta_i^{(X)} = -X_{i1}+0.5X_{i2}-0.25X_{i3}-0.1X_{i4} $, while $ \delta_T = U $ implies a nonlinear relaltionship between the log odds of the propensity score and $ \eta_i^{(U)} = -U_{i1}+0.5U_{i2}-0.25U_{i3}-0.1U_{i4}$. As for the outcome propcess, we know that $ Y = Y(1) \cdot T + Y(0) \cdot (1-T)$ and construct the pair potential outcomes $ \{Y_i(1),Y_i(0)\} $ for unit $ i,~i=1,\cdots,N $ with a bivariate model
   \[
   \begin{pmatrix}
   Y_i(0)\\Y_i(1)
   \end{pmatrix}^{(\delta_O)} \sim N\begin{pmatrix}
   \left[ \begin{array}{c}
   \mu_i^{(\delta_O)}\\\mu_i^{(\delta_O)}+\gamma
   \end{array} 
   \right ] , \left[ \begin{array}{cc}
   \sigma^2 & \rho \sigma \\
   \rho \sigma & \sigma^2\\
   \end{array} 
   \right] 
   \end{pmatrix} 
   \]
where $ \sigma^2 $ is the error variance, $ \rho $ is the correlation between the potential outcomes, $ \gamma = 20 $ represents the treatment effect and $ \delta_O \in \{X,U\} $ indicates the covariates, whether $ \bm{X} $ or $ \bm{U} $, that participate in the outcome process. For example, if $ \delta_O = X $, we have $ \mu_i^{(X)} = 210+27.4X_{i1}+13.7X_{i2}+13.7X_{i3}+13.7X_{i4} $. It should be emphasized that although we may use $ \bm{U} $ to construct variable $ T $ or $ Y $, it cannot be observed and the covariates to be balanced is always $ \bm{X} $. In the other words, the observed simulated data consists of $ \{X_i, T_i,Y_i \}, i =1,\cdots,N$. 

Based on this data generation mechanism, we can know that 
\begin{itemize}
	\item ATE:
	$$ \begin{aligned}
	\tau &= E\left[ Y_i(1) - Y_i(0) \right] = E\left[ Y_i(1)\right] - E\left[Y_i(0) \right]  \\
	&= \left( \mu_i^{(\delta_O)}+\gamma \right) - \mu_i^{(\delta_O)} = \gamma = 20
	\end{aligned} $$ 
	
	\item ATT:
	$$ \begin{aligned}
	\tau_{trt} &= E\left[ Y_i(1) - Y_i(0) |T=1 \right] \\
	&= E\left[ \left( \mu_i^{(\delta_O)}+\gamma \right) - \mu_i^{(\delta_O)}  |T=1 \right] \\
	&= \gamma = 20
	\end{aligned} $$ 
\end{itemize}

To better understand the role of different parameters on estimating ATE, we choose the sample size $ N $ in $ \{200,1000\} $, the error variance $ \sigma^2 $ in $ \{ 2,5,10 \}$, the correlation between the potential outcomes $\rho $ in $ \{ -0.3,0,0.5 \}$, the generative process $ \delta_T $ that determines the treatment assignment in \{$ U $: $ T $ is generated with covariates $ \bm{U} $; $ X $: $ T $ is generated with covariates $ \bm{X} $ \}, and the outcome process $ \delta_O $ in \{$ U $: $ Y $ is generated with covariates $ \bm{U} $; $ X $: $ T $ is generated with covariates $ \bm{X} $\}, so there are a total of 72 different simulaiton scenarios. We change only one parameter in different simulation scenarios and run 500 Monte Carlo datasets for each setting.
 
\begin{figure}[H]
 	\centering
 	\includegraphics[width=\textwidth,height=0.7\textheight]{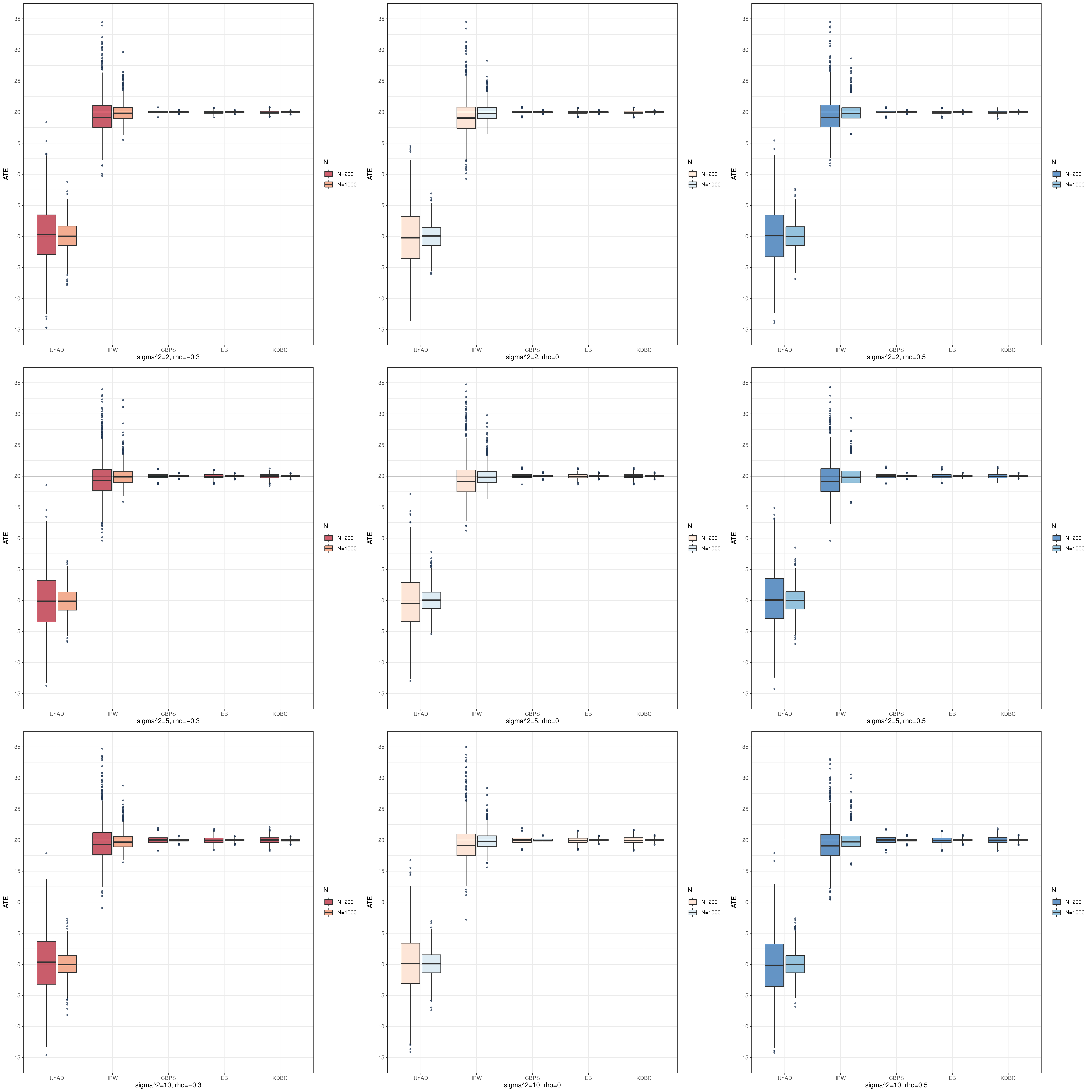}
 	\caption{T Scenario = ``X'', Y Scenario = ``X''}
 	\label{Sig_rho_N_comparison_XX}
 \end{figure}

\begin{figure}[H]
	\centering
	\includegraphics[width=\textwidth,height=0.7\textheight]{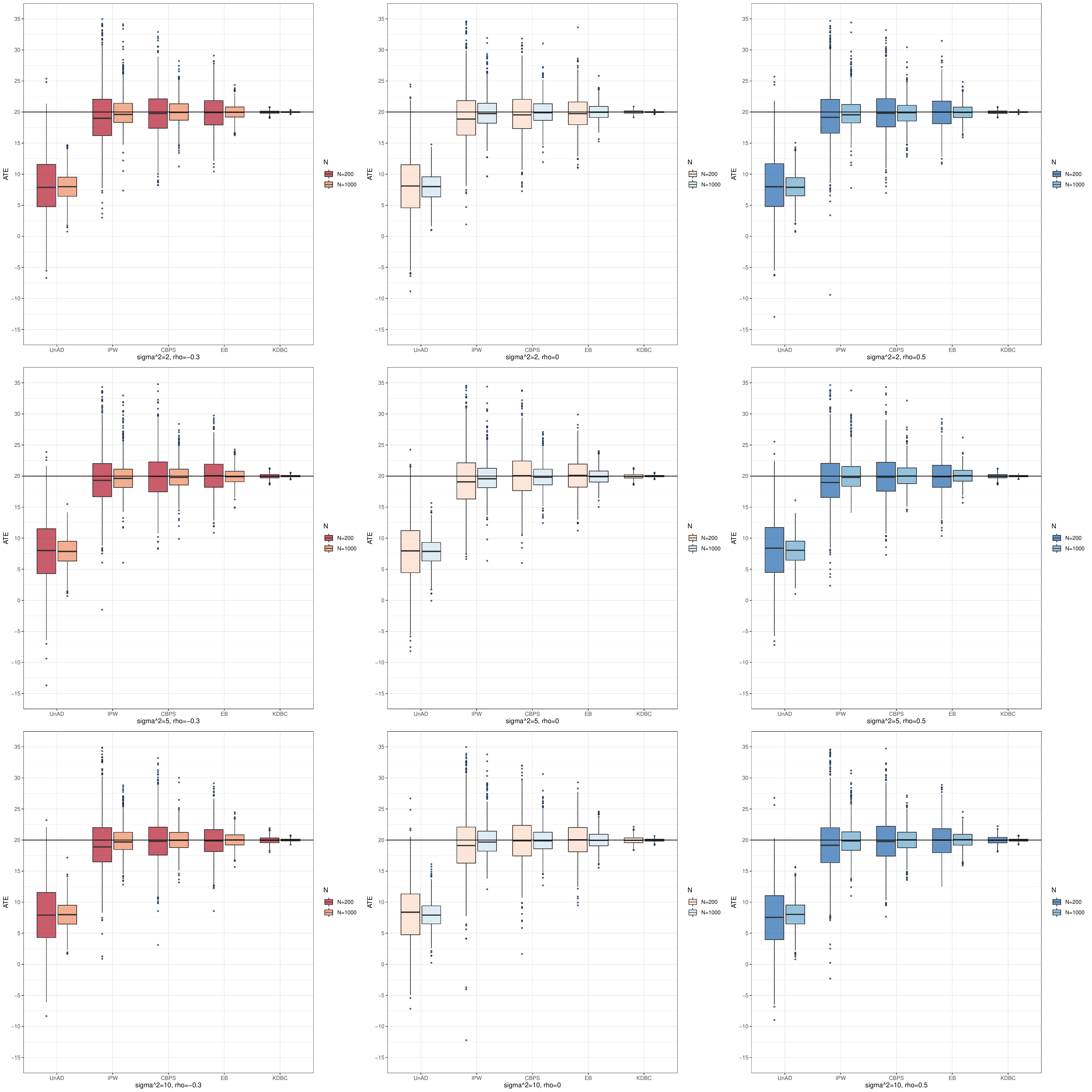}
	\caption{T Scenario = ``X'', Y Scenario = ``U''}
	\label{Sig_rho_N_comparison_XU}
\end{figure}

\begin{figure}[H]
	\centering
	\includegraphics[width=\textwidth,height=0.7\textheight]{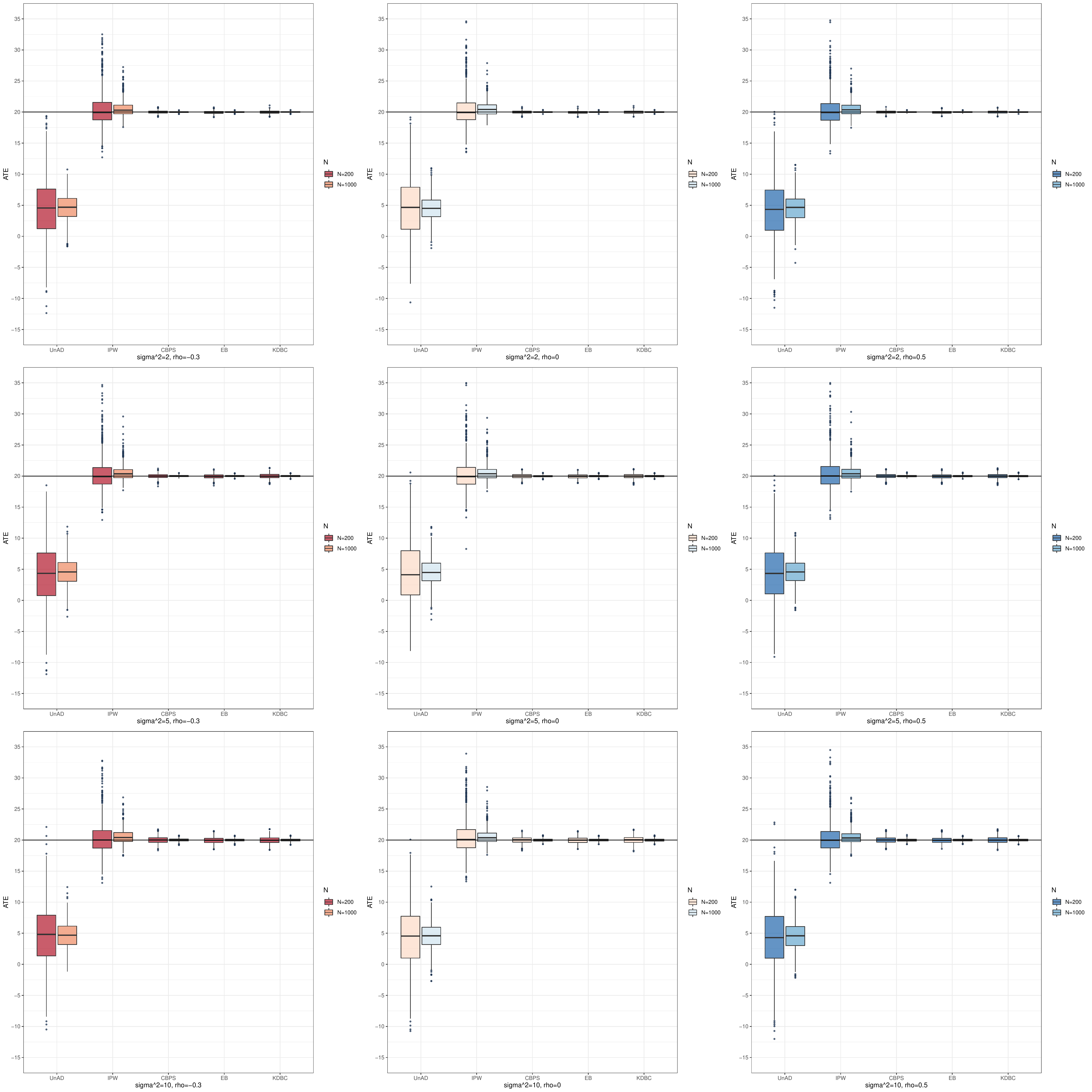}
	\caption{T Scenario = ``U'', Y Scenario = ``X''.}
	\label{Sig_rho_N_comparison_UX}
\end{figure}

\begin{figure}[H]
\centering
\includegraphics[width=\textwidth,height=0.7\textheight]{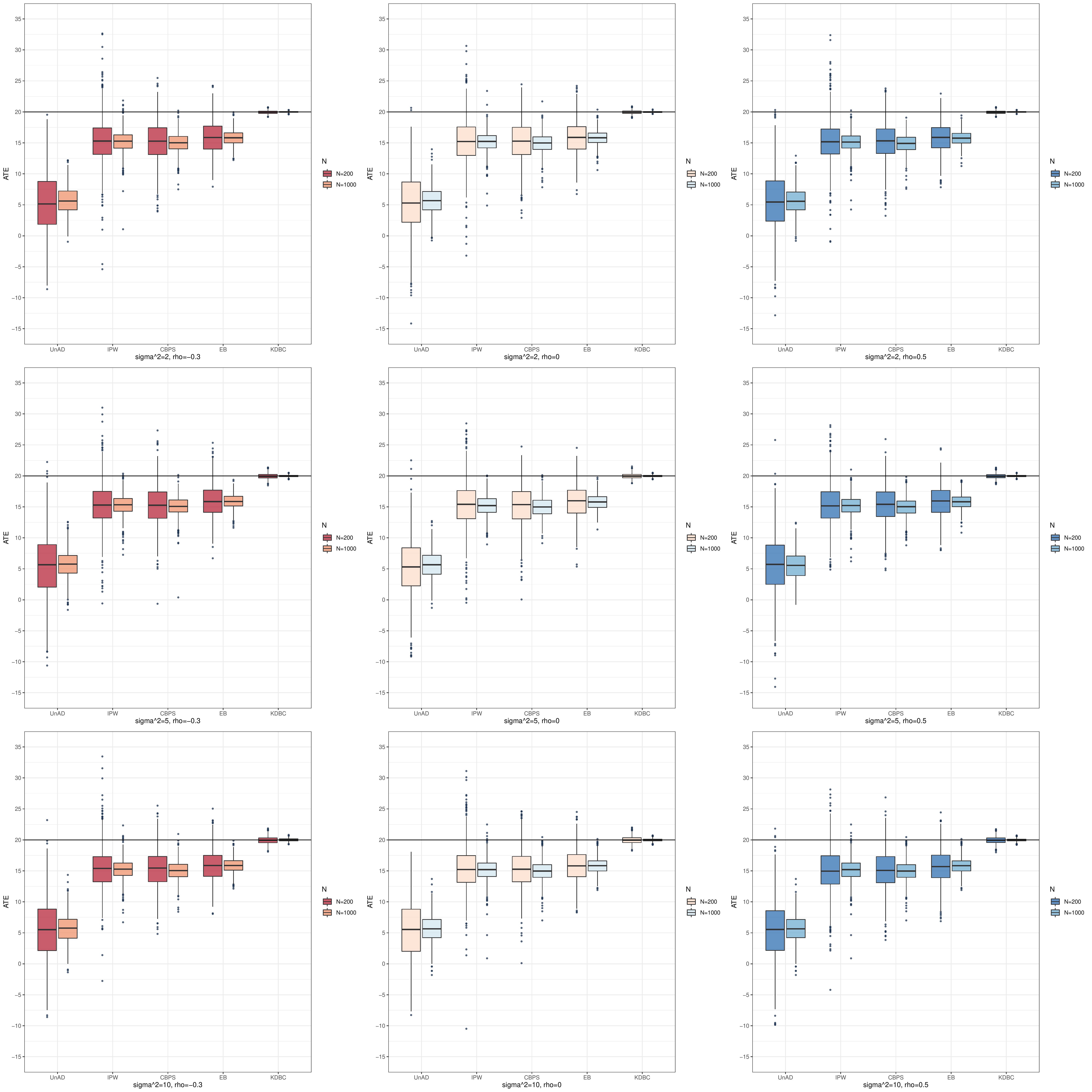}
\caption{T Scenario = ``U'', Y Scenario = ``U''}
\label{Sig_rho_N_comparison_UU}
\end{figure}

Fixing $ \delta_T $ and $ \delta_O $ (\{$ \delta_T $ = X, $ \delta_O $ = X\},  \{$ \delta_T $ = X, $ \delta_O $ = U\}, \{$ \delta_O $ = U, $ \delta_O $ = X\}, \{$ \delta_T $ = U, $ \delta_O $ = U\}, we run 1000 Monte Carlo datasets to compare the performance of four weighting methods (IPW, CBPS, EB, KDBC) on estimating ATE under different combinations of $N$, $\sigma^2$ and $\rho$. The results are briefly shown in Figure [\ref{Sig_rho_N_comparison_XX},
 \ref{Sig_rho_N_comparison_XU}, \ref{Sig_rho_N_comparison_UX}, \ref{Sig_rho_N_comparison_UU}]. For every Figure corresponds to a different combination of $ \delta_T $ and $ \delta_O $, the value of $\sigma^2$ corresponds to each row from top to bottom are $\{2,5,10\}$, and the  value of $\rho$ corresponds to each column from left to right are $\{-0.3,0,0.5\}$. It can be found that the sample size $N$ and the error variance $ \sigma^2$ mainly influence the standard errors of ATE estimators, while the correlation between potential outcomes $Y_i(0)$ and $Y_i(1)$ has almost no effect. 
 This observation is consistent with \citet{josey2020framework} and thus, we focus on the simulation scenario with parameters $N = 200$, $\sigma^2 = 10$, and $\rho = 0$. 
 
\begin{figure}[H]
	\centering
	\includegraphics[width=\textwidth]{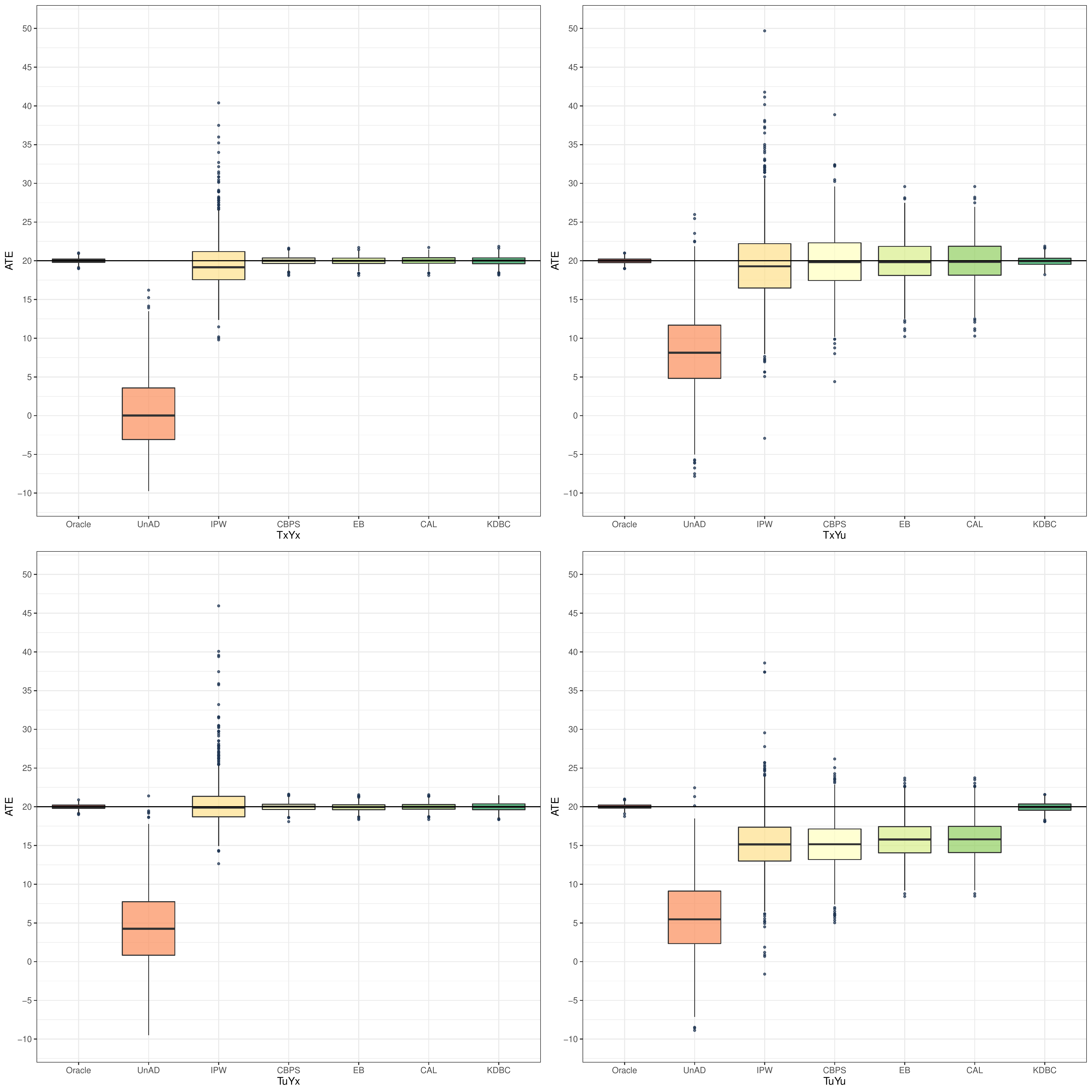}
	\caption{\textbf{ATE} estimate using five balancing methods (\textbf{IPW}, \textbf{CBPS}, \textbf{EB}, \textbf{CAL} and \textbf{KDBC}) with $N = 200$, $\sigma^2 = 10$, $\rho = 0$. Both oracle estimate and unadjusted (\textbf{UnAD}) estimate are used as a benchmark for comparison.}
	\label{N200Sig10rho0ATE2x2_X}
\end{figure}

We run 500 Monte Carlo experiments for estimating both the ATE and ATT. The results of ATE estimation are shown in Figure [\ref{N200Sig10rho0ATE2x2_X}] and Table [\ref{Table4ATEstimate_X},\ref{Table4BalanceMetric_X}], and the results of ATT estimation are shown in Table [\ref{Table4ATTEstimate_X},\ref{Table4ATTBalanceMetric_X}]. Table [\ref{Table4ATEstimate_X},\ref{Table4ATTEstimate_X}] together show that the KDBC estimators, both for ATE and ATT, can perform as well as the other weighting estimators when all the true predictors are included in the observed dataset. While in both scenarios of \{$ \delta_T $ = X, $ \delta_O $ = U\} and \{$ \delta_T $ = U, $ \delta_O $ = U\} where the key information of the true covariates are missed, the KDBC estimators can perform much better than other weighting estimators. This reflects the high stability of our proposed KDB covariate balance in estimating causal effects even when the observed dataset misses some key information about the real generation mechanism.
 
\begin{figure}[H]
	\centering
	\includegraphics[width=\textwidth,height=0.95\textheight]{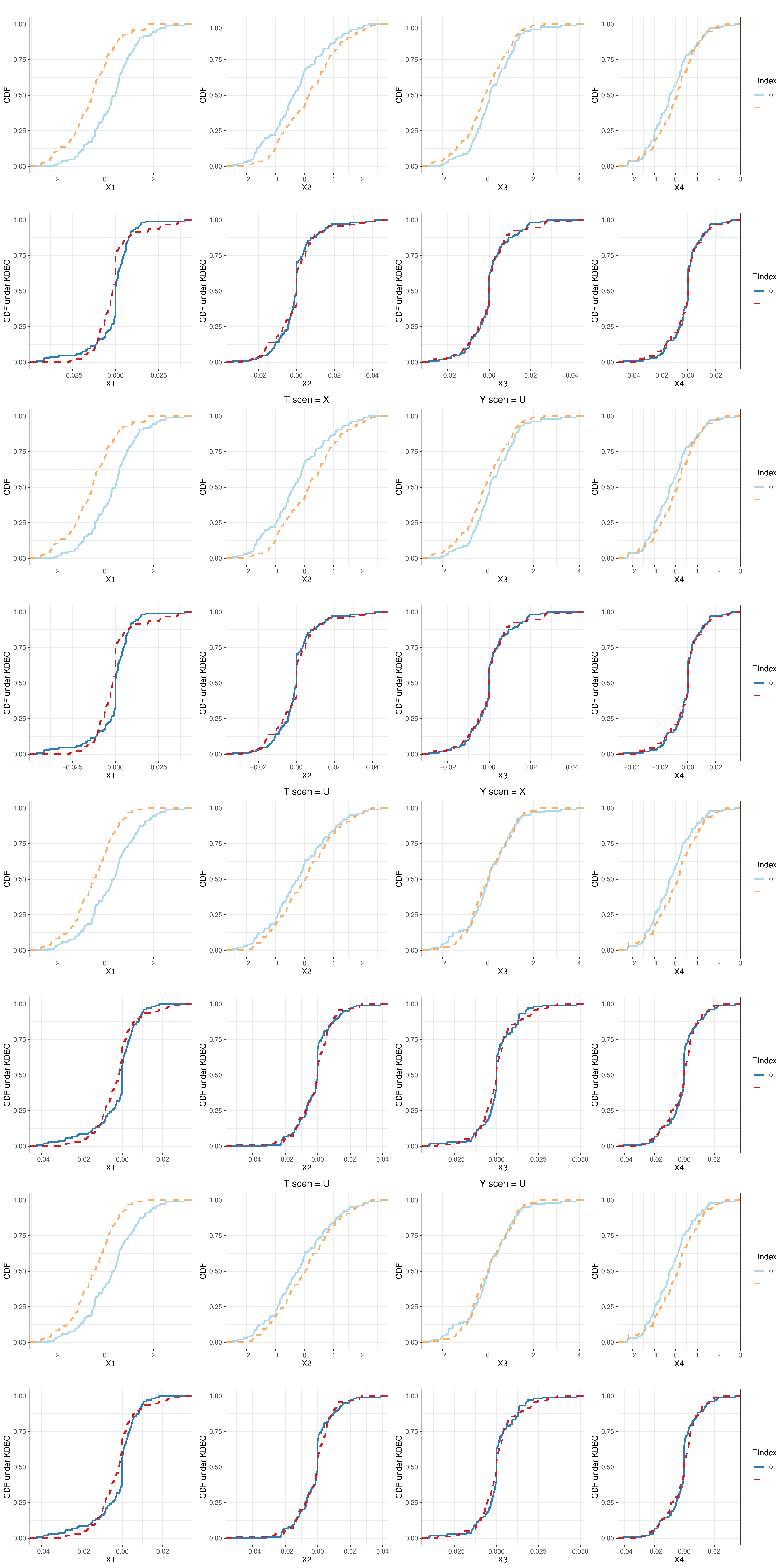}
	\caption{The CDF of unweighted covariates is arranged in odd rows, and that of weighted covariates under KDBC is arranged in even rows. All combination of ``T scenario'' and ``Y scenario'' are included. Dashed line indicates the treatment group, and the solid line indicatesv the control group.}
	\label{Sim1BalancingCDF_AllTAllY_X}
\end{figure}

To better observe the adjustment effect of our proposed method, we draw the cumulative distribution functions (CDF) of all covariates without any adjustment and that of the weighted covariates under KDBC. It can be seen from Figure [\ref{Sim1BalancingCDF_AllTAllY_X}] that regardless of the value of $ \delta_T $ and $ \delta_O $, under our proposed KDB covariate balancing with only regularization constraints (KDBC), the CDFs of the weighted covariates between the treatment and control groups can always match each other well. This means that our proposed method can effectively achieve the covariate balance between the treatment and control groups.
 
\begin{table}[H]
	\centering
	\resizebox{\linewidth}{!}{
		\begin{threeparttable}
			\begin{tabular}{ccccccccccccccccccccc}
				\toprule
				\multicolumn{1}{c}{\multirow{2}[4]{*}{\textbf{\tabincell{c}{Treatment\\ Assignment\\ Scenario}}}} & \multirow{2}[4]{*}{} & \multicolumn{1}{c}{\multirow{2}[4]{*}{\textbf{\tabincell{c}{Outcome \\Scenario}}}} & \multirow{2}[4]{*}{} & \multirow{2}[4]{*}{\textbf{Metric}} & \multirow{2}[4]{*}{} & \multicolumn{3}{c}{\textbf{Benchmark}} & \multirow{2}[4]{*}{} & \multicolumn{11}{c}{\textbf{Balancing Methods}} \\
				\cmidrule{7-9}\cmidrule{11-21}          &       &       &       &       &       & \textbf{Oracle} &       & \textbf{UnAD} &       & \textbf{IPW} &       & \textbf{CBPS} &       & \textbf{EB} &       & \textbf{CAL} &       & \textbf{KDBC} &       & \textbf{KDM1} \\
				\midrule
				\multirow{4}[2]{*}{X} &       & \multirow{4}[2]{*}{X} &       & \textbf{ATE} &       & 20.02181  &       & -0.22254  &       & 19.42157  &       & 19.99264  &       & 19.97454  &       & 20.00381  &       & 19.99452  &       & 20.01701  \\
				&       &       &       & \textbf{Bias} &       & 0.02181  &       & -20.22254  &       & -0.57843  &       & -0.00736  &       & -0.02546  &       & 0.00381  &       & -0.00548  &       & 0.01701  \\
				&       &       &       & \textbf{sd} &       & 0.01358  &       & 0.22983  &       & 0.16986  &       & 0.02390  &       & 0.02298  &       & 0.02302  &       & 0.03207  &       & 0.03171  \\
				&       &       &       & \textbf{RMSE} &       & 0.01360  &       & 0.93307  &       & 0.17165  &       & 0.02388  &       & 0.02299  &       & 0.02299  &       & 0.03204  &       & 0.03168  \\
				\midrule
				\midrule
				\multirow{4}[2]{*}{X} &       & \multirow{4}[2]{*}{U} &       & \textbf{ATE} &       & 20.01750  &       & 7.76609  &       & 19.28762  &       & 19.67198  &       & 19.74617  &       & 19.76634  &       & 19.95198  &       & 19.97879  \\
				&       &       &       & \textbf{Bias} &       & 0.01750  &       & -12.23391  &       & -0.71238  &       & -0.32802  &       & -0.25383  &       & -0.23366  &       & -0.04802  &       & -0.02121  \\
				&       &       &       & \textbf{sd} &       & 0.01428  &       & 0.23171  &       & 0.25003  &       & 0.17427  &       & 0.12919  &       & 0.12925  &       & 0.03289  &       & 0.03279  \\
				&       &       &       & \textbf{RMSE} &       & 0.01428  &       & 0.59407  &       & 0.25181  &       & 0.17471  &       & 0.12956  &       & 0.12954  &       & 0.03293  &       & 0.03277  \\
				\midrule
				\midrule
				\multirow{4}[2]{*}{U} &       & \multirow{4}[2]{*}{X} &       & \textbf{ATE} &       & 19.99365  &       & 4.149422956 &       & 20.37174  &       & 19.99844  &       & 19.95353  &       & 19.99821  &       & 19.98244  &       & 19.99608  \\
				&       &       &       & \textbf{Bias} &       & -0.00635  &       & -15.85057704 &       & 0.37174  &       & -0.00156  &       & -0.04647  &       & -0.00179  &       & -0.01756  &       & -0.00392  \\
				&       &       &       & \textbf{sd} &       & 0.01431  &       & 0.215249227 &       & 0.14953  &       & 0.02376  &       & 0.02326  &       & 0.02298  &       & 0.03060  &       & 0.03047  \\
				&       &       &       & \textbf{RMSE} &       & 0.01429  &       & 0.740757147 &       & 0.15030  &       & 0.02374  &       & 0.02333  &       & 0.02295  &       & 0.03058  &       & 0.03044  \\
				\midrule
				\midrule
				\multirow{4}[2]{*}{U} &       & \multirow{4}[2]{*}{U} &       & \textbf{ATE} &       & 19.99174  &       & 5.308382057 &       & 15.30943  &       & 15.19822  &       & 15.83007  &       & 15.86088  &       & 19.89995  &       & 19.91026  \\
				&       &       &       & \textbf{Bias} &       & -0.00826  &       & -14.69161794 &       & -4.69057  &       & -4.80178  &       & -4.16993  &       & -4.13912  &       & -0.10005  &       & -0.08974  \\
				&       &       &       & \textbf{sd} &       & 0.01474  &       & 0.223602406 &       & 0.17365  &       & 0.14242  &       & 0.11789  &       & 0.11796  &       & 0.03044  &       & 0.03037  \\
				&       &       &       & \textbf{RMSE} &       & 0.01473  &       & 0.69396  &       & 0.27221  &       & 0.25760  &       & 0.22056  &       & 0.21943  &       & 0.03074  &       & 0.03061  \\
				\bottomrule
			\end{tabular}%
		\end{threeparttable}
	}
	\caption{Details on ATE estimate with 500 Monte Carlo datasets.}
	\label{Table4ATEstimate_X}%
\end{table}%

\begin{table}[H]
	\centering
	\resizebox{\linewidth}{!}{
		\begin{threeparttable}
			\begin{tabular}{cccrcrccccccccccccccc}
				\toprule
				\multicolumn{1}{c}{\multirow{2}[4]{*}{\textbf{\tabincell{c}{Treatment\\ Assignment\\ Scenario}}}} & \multirow{2}[4]{*}{} & \multicolumn{1}{c}{\multirow{2}[4]{*}{\textbf{\tabincell{c}{Outcome \\Scenario}}}} & \multirow{2}[4]{*}{} & \multirow{2}[4]{*}{\textbf{Balancing Metric}} & \multirow{2}[4]{*}{} & \multicolumn{3}{c}{\textbf{Benchmark}} & \multirow{2}[4]{*}{} & \multicolumn{11}{c}{\textbf{Balancing Methods}} \\
				\cmidrule{7-9}\cmidrule{11-21}          &       &       &       &       &       & \textbf{Oracle} &       & \textbf{UnAD} &       & \textbf{IPW} &       & \textbf{CBPS} &       & \textbf{EB} &       & \textbf{CAL} &       & \textbf{KDBC} &       & \textbf{KDM1} \\
				\midrule
				\multirow{6}[2]{*}{X} &       & \multirow{6}[2]{*}{X} &       & \textbf{KD} &       & 0.00100  &       & 0.17749  &       & 0.03689  &       & 0.02457  &       & 0.02138  &       & 0.02140  &       & 0.00036  &       & 0.00035  \\
				&       &       &       & \textbf{maxASMD} &       & 0.00000  &       & 0.00879  &       & 0.00129  &       & 0.00015  &       & 0.00012  &       & 0.00012  &       & 0.00017  &       & 0.00017  \\
				&       &       &       & \textbf{meanASMD} &       & 0.00000  &       & 0.00409  &       & 0.00068  &       & 0.00008  &       & 0.00007  &       & 0.00007  &       & 0.00009  &       & 0.00009  \\
				&       &       &       & \textbf{medASMD} &       & 0.00000  &       & 0.00327  &       & 0.00061  &       & 0.00007  &       & 0.00006  &       & 0.00006  &       & 0.00008  &       & 0.00008  \\
				&       &       &       & \textbf{meanKS} &       & 0.00000  &       & 0.22341  &       & 0.21086  &       & 0.21025  &       & 0.20455  &       & 0.20455  &       & 0.20812  &       & 0.20635  \\
				&       &       &       & \textbf{meanT} &       & 0.00000  &       & -1.30843  &       & -0.05517  &       & 0.00350  &       & 0.00279  &       & 0.00439  &       & -0.00104  &       & -0.00060  \\
				\midrule
				\midrule
				\multirow{6}[2]{*}{X} &       & \multirow{6}[2]{*}{U} &       & \textbf{KD} &       & 0.00100  &       & 0.17657  &       & 0.03709  &       & 0.02459  &       & 0.02145  &       & 0.02147  &       & 0.00036  &       & 0.00036  \\
				&       &       &       & \textbf{maxASMD} &       & 0.00000  &       & 0.00874  &       & 0.00129  &       & 0.00016  &       & 0.00013  &       & 0.00013  &       & 0.00018  &       & 0.00018  \\
				&       &       &       & \textbf{meanASMD} &       & 0.00000  &       & 0.00404  &       & 0.00069  &       & 0.00009  &       & 0.00007  &       & 0.00007  &       & 0.00010  &       & 0.00010  \\
				&       &       &       & \textbf{medASMD} &       & 0.00000  &       & 0.00324  &       & 0.00062  &       & 0.00008  &       & 0.00006  &       & 0.00006  &       & 0.00009  &       & 0.00009  \\
				&       &       &       & \textbf{meanKS} &       & 0.00000  &       & 0.22388  &       & 0.21115  &       & 0.21048  &       & 0.20520  &       & 0.20520  &       & 0.20888  &       & 0.20717  \\
				&       &       &       & \textbf{meanT} &       & 0.00000  &       & -1.28633  &       & -0.03970  &       & 0.00588  &       & 0.00577  &       & 0.00748  &       & 0.00066  &       & 0.00144  \\
				\midrule
				\midrule
				\multirow{6}[2]{*}{U} &       & \multirow{6}[2]{*}{X} &       & \textbf{KD} &       & 0.00100  &       & 0.15692  &       & 0.03542  &       & 0.02824  &       & 0.02555  &       & 0.02557  &       & 0.00036  &       & 0.00035  \\
				&       &       &       & \textbf{maxASMD} &       & 0.00000  &       & 0.00837  &       & 0.00101  &       & 0.00014  &       & 0.00012  &       & 0.00011  &       & 0.00016  &       & 0.00016  \\
				&       &       &       & \textbf{meanASMD} &       & 0.00000  &       & 0.00359  &       & 0.00053  &       & 0.00008  &       & 0.00006  &       & 0.00006  &       & 0.00009  &       & 0.00009  \\
				&       &       &       & \textbf{medASMD} &       & 0.00000  &       & 0.00265  &       & 0.00047  &       & 0.00007  &       & 0.00006  &       & 0.00005  &       & 0.00008  &       & 0.00008  \\
				&       &       &       & \textbf{meanKS} &       & 0.00000  &       & 0.20399  &       & 0.19083  &       & 0.19077  &       & 0.18849  &       & 0.18850  &       & 0.19127  &       & 0.19017  \\
				&       &       &       & \textbf{meanT} &       & 0.00000  &       & -0.85087  &       & 0.02321  &       & 0.00628  &       & -0.00105  &       & 0.00097  &       & 0.00171  &       & 0.00224  \\
				\midrule
				\midrule
				\multirow{6}[2]{*}{U} &       & \multirow{6}[2]{*}{U} &       & \textbf{KD} &       & 0.00100  &       & 0.15523  &       & 0.03651  &       & 0.02824  &       & 0.02557  &       & 0.02558  &       & 0.00035  &       & 0.00036  \\
				&       &       &       & \textbf{maxASMD} &       & 0.00000  &       & 0.00834  &       & 0.00109  &       & 0.00015  &       & 0.00012  &       & 0.00012  &       & 0.00017  &       & 0.00017  \\
				&       &       &       & \textbf{meanASMD} &       & 0.00000  &       & 0.00354  &       & 0.00057  &       & 0.00008  &       & 0.00007  &       & 0.00006  &       & 0.00009  &       & 0.00009  \\
				&       &       &       & \textbf{medASMD} &       & 0.00000  &       & 0.00258  &       & 0.00050  &       & 0.00007  &       & 0.00006  &       & 0.00006  &       & 0.00009  &       & 0.00009  \\
				&       &       &       & \textbf{meanKS} &       & 0.00000  &       & 0.20276  &       & 0.18890  &       & 0.18875  &       & 0.18557  &       & 0.18556  &       & 0.18917  &       & 0.18809  \\
				&       &       &       & \textbf{meanT} &       & 0.00000  &       & -0.85721  &       & 0.03375  &       & 0.00702  &       & -0.00030  &       & 0.00165  &       & 0.00152  &       & 0.00210  \\
				\bottomrule
			\end{tabular}%
		\end{threeparttable}
	}
	\caption{ATE - Details on Balancing Metrics with 500 Monte Carlo datasets.}
	\label{Table4BalanceMetric_X}%
\end{table}%

\begin{table}[H]
	\centering
	\resizebox{\linewidth}{!}{
		\begin{threeparttable}
			\begin{tabular}{cccccccrcrcrcrcrcrc}
				\toprule	\multicolumn{1}{c}{\multirow{3}[4]{*}{\textbf{\tabincell{c}{Treatment\\ Assignment\\ Scenario}}}} & \multirow{3}[4]{*}{} & \multicolumn{1}{c}{\multirow{3}[4]{*}{\textbf{\tabincell{c}{Outcome\\ Scenario}}}} & \multirow{3}[4]{*}{} & \multirow{3}[4]{*}{\textbf{Metric}} & \multirow{3}[4]{*}{} & \multirow{2}[2]{*}{\textbf{Benchmark}} & \multirow{3}[4]{*}{} & \multicolumn{11}{c}{\multirow{2}[2]{*}{\textbf{Balancing Methods}}} \\
				&       &       &       &       &       &       &       & \multicolumn{11}{c}{} \\
				\cmidrule{7-7}\cmidrule{9-19}          &       &       &       &       &       & \textbf{Oracle} &       & \textbf{IPW} &       & \textbf{CBPS} &       & \textbf{EB} &       & \textbf{CAL} &       & \textbf{KDM1} &       & \textbf{KDBC} \\
				\midrule
				\multirow{4}[2]{*}{X} &       & \multirow{4}[2]{*}{X} &       & \textbf{ATT} &       & 20.02181  &       & 18.94827  &       & 20.01480  &       & 19.94777  &       & 20.01839  &       & 20.01063  &       & 19.48173  \\
				&       &       &       & \textbf{Bias} &       & 0.02181  &       & -1.05173  &       & 0.01480  &       & -0.05223  &       & 0.01839  &       & 0.01063  &       & -0.51827  \\
				&       &       &       & \textbf{sd} &       & 0.01358  &       & 0.22240  &       & 0.02916  &       & 0.02931  &       & 0.02924  &       & 0.04480  &       & 0.04704  \\
				&       &       &       & \textbf{RMSE} &       & 0.01360  &       & 0.22710  &       & 0.02914  &       & 0.02938  &       & 0.02922  &       & 0.04476  &       & 0.05240  \\
				\midrule
				\midrule
				\multirow{4}[2]{*}{X} &       & \multirow{4}[2]{*}{U} &       & \textbf{ATT} &       & 20.01750  &       & 19.69548  &       & 20.77572  &       & 20.75489  &       & 20.77844  &       & 20.41813  &       & 19.82211  \\
				&       &       &       & \textbf{Bias} &       & 0.01750  &       & -0.30452  &       & 0.77572  &       & 0.75489  &       & 0.77844  &       & 0.41813  &       & -0.17789  \\
				&       &       &       & \textbf{sd} &       & 0.01428  &       & 0.24458  &       & 0.15395  &       & 0.15362  &       & 0.15389  &       & 0.05750  &       & 0.05277  \\
				&       &       &       & \textbf{RMSE} &       & 0.01428  &       & 0.24471  &       & 0.15766  &       & 0.15714  &       & 0.15763  &       & 0.06041  &       & 0.05332  \\
				\midrule
				\midrule
				\multirow{4}[2]{*}{U} &       & \multirow{4}[2]{*}{X} &       & \textbf{ATT} &       & 19.99365  &       & 25.08391  &       & 19.99283  &       & 19.94235  &       & 19.99311  &       & 19.97513  &       & 19.77988  \\
				&       &       &       & \textbf{Bias} &       & -0.00635  &       & 5.08391  &       & -0.00717  &       & -0.05765  &       & -0.00689  &       & -0.02487  &       & -0.22012  \\
				&       &       &       & \textbf{sd} &       & 0.01431  &       & 0.25533  &       & 0.02654  &       & 0.02661  &       & 0.02653  &       & 0.04041  &       & 0.04144  \\
				&       &       &       & \textbf{RMSE} &       & 0.01429  &       & 0.34170  &       & 0.02652  &       & 0.02671  &       & 0.02650  &       & 0.04038  &       & 0.04255  \\
				\midrule
				\midrule
				\multirow{4}[2]{*}{U} &       & \multirow{4}[2]{*}{U} &       & \textbf{ATT} &       & 19.99174  &       & 19.07117  &       & 16.39781  &       & 16.37561  &       & 16.39804  &       & 19.96664  &       & 19.74746  \\
				&       &       &       & \textbf{Bias} &       & -0.00826  &       & -0.92883  &       & -3.60219  &       & -3.62439  &       & -3.60196  &       & -0.03336  &       & -0.25254  \\
				&       &       &       & \textbf{sd} &       & 0.01474  &       & 0.25013  &       & 0.15951  &       & 0.15934  &       & 0.15953  &       & 0.04586  &       & 0.04628  \\
				&       &       &       & \textbf{RMSE} &       & 0.01473  &       & 0.25331  &       & 0.22659  &       & 0.22718  &       & 0.22660  &       & 0.04584  &       & 0.04759  \\
				\bottomrule
			\end{tabular}%
		\end{threeparttable}
	}
	\caption{Details on ATT estimate with 500 Monte Carlo datasets.}
	\label{Table4ATTEstimate_X}%
\end{table}

\begin{table}[H]
	\centering
	\resizebox{\linewidth}{!}{
		\begin{threeparttable}
			\begin{tabular}{cccrcrcrcrcrcrcrcrc}
				\toprule
				\multicolumn{1}{c}{\multirow{3}[4]{*}{\textbf{\tabincell{c}{Treatment\\ Assignment\\ Scenario}}}} & \multirow{3}[4]{*}{} & \multicolumn{1}{c}{\multirow{3}[4]{*}{\textbf{\tabincell{c}{Outcome\\ Scenario}}}} & \multirow{3}[4]{*}{} & \multirow{3}[4]{*}{\textbf{\tabincell{c}{Balancing\\ Metric}}} & \multirow{3}[4]{*}{} & \multirow{2}[2]{*}{\textbf{Benchmark}} & \multirow{3}[4]{*}{} & \multicolumn{11}{c}{\multirow{2}[2]{*}{\textbf{Balancing Methods}}} \\
				&       &       &       &       &       &       &       & \multicolumn{11}{c}{} \\
				\cmidrule{7-7}\cmidrule{9-19}          &       &       &       &       &       & \textbf{Oracle} &       & \textbf{IPW} &       & \textbf{CBPS} &       & \textbf{EB} &       & \textbf{CAL} &       & \textbf{KDM1} &       & \textbf{KDBC} \\
				\midrule
				\multirow{6}[2]{*}{X} &       & \multirow{6}[2]{*}{X} &       & \textbf{KD} &       & 0.00142  &       & 0.05126  &       & 0.02618  &       & 0.02612  &       & 0.02619  &       & 0.00725  &       & 0.00568  \\
				&       &       &       & \textbf{maxASMD} &       & 0.00000  &       & 0.00198  &       & 0.00050  &       & 0.00050  &       & 0.00050  &       & 0.00050  &       & 0.00054  \\
				&       &       &       & \textbf{meanASMD} &       & 0.00000  &       & 0.00107  &       & 0.00024  &       & 0.00024  &       & 0.00024  &       & 0.00024  &       & 0.00027  \\
				&       &       &       & \textbf{medASMD} &       & 0.00000  &       & 0.00098  &       & 0.00020  &       & 0.00020  &       & 0.00020  &       & 0.00020  &       & 0.00023  \\
				&       &       &       & \textbf{meanKS} &       & 0.00000  &       & 0.25172  &       & 0.26461  &       & 0.26408  &       & 0.26467  &       & 0.45139  &       & 0.43743  \\
				&       &       &       & \textbf{meanT} &       & 0.00000  &       & -0.09005  &       & -0.00207  &       & -0.00467  &       & -0.00190  &       & -0.00225  &       & -0.01696  \\
				\midrule
				\midrule
				\multirow{6}[2]{*}{X} &       & \multirow{6}[2]{*}{U} &       & \textbf{KD} &       & 0.00142  &       & 0.05214  &       & 0.02675  &       & 0.02669  &       & 0.02676  &       & 0.00713  &       & 0.00556  \\
				&       &       &       & \textbf{maxASMD} &       & 0.00000  &       & 0.00205  &       & 0.00051  &       & 0.00051  &       & 0.00051  &       & 0.00051  &       & 0.00055  \\
				&       &       &       & \textbf{meanASMD} &       & 0.00000  &       & 0.00110  &       & 0.00024  &       & 0.00024  &       & 0.00024  &       & 0.00024  &       & 0.00027  \\
				&       &       &       & \textbf{medASMD} &       & 0.00000  &       & 0.00100  &       & 0.00020  &       & 0.00020  &       & 0.00020  &       & 0.00020  &       & 0.00022  \\
				&       &       &       & \textbf{meanKS} &       & 0.00000  &       & 0.25165  &       & 0.26293  &       & 0.26249  &       & 0.26299  &       & 0.44922  &       & 0.43451  \\
				&       &       &       & \textbf{meanT} &       & 0.00000  &       & -0.05648  &       & -0.00233  &       & -0.00499  &       & -0.00215  &       & -0.00118  &       & -0.01506  \\
				\midrule
				\midrule
				\multirow{6}[2]{*}{U} &       & \multirow{6}[2]{*}{X} &       & \textbf{KD} &       & 0.00140  &       & 0.05612  &       & 0.02820  &       & 0.02814  &       & 0.02820  &       & 0.00501  &       & 0.00412  \\
				&       &       &       & \textbf{maxASMD} &       & 0.00000  &       & 0.00232  &       & 0.00041  &       & 0.00040  &       & 0.00041  &       & 0.00041  &       & 0.00041  \\
				&       &       &       & \textbf{meanASMD} &       & 0.00000  &       & 0.00123  &       & 0.00018  &       & 0.00018  &       & 0.00018  &       & 0.00018  &       & 0.00019  \\
				&       &       &       & \textbf{medASMD} &       & 0.00000  &       & 0.00110  &       & 0.00014  &       & 0.00014  &       & 0.00014  &       & 0.00014  &       & 0.00016  \\
				&       &       &       & \textbf{meanKS} &       & 0.00000  &       & 0.22078  &       & 0.21968  &       & 0.21934  &       & 0.21968  &       & 0.41261  &       & 0.39862  \\
				&       &       &       & \textbf{meanT} &       & 0.00000  &       & 0.30059  &       & 0.00518  &       & 0.00423  &       & 0.00518  &       & 0.00598  &       & 0.00305  \\
				\midrule
				\midrule
				\multirow{6}[2]{*}{U} &       & \multirow{6}[2]{*}{U} &       & \textbf{KD} &       & 0.00141  &       & 0.05644  &       & 0.02832  &       & 0.02827  &       & 0.02832  &       & 0.00489  &       & 0.00403  \\
				&       &       &       & \textbf{maxASMD} &       & 0.00000  &       & 0.00237  &       & 0.00044  &       & 0.00044  &       & 0.00044  &       & 0.00044  &       & 0.00044  \\
				&       &       &       & \textbf{meanASMD} &       & 0.00000  &       & 0.00123  &       & 0.00020  &       & 0.00020  &       & 0.00020  &       & 0.00020  &       & 0.00021  \\
				&       &       &       & \textbf{medASMD} &       & 0.00000  &       & 0.00108  &       & 0.00015  &       & 0.00015  &       & 0.00015  &       & 0.00015  &       & 0.00017  \\
				&       &       &       & \textbf{meanKS} &       & 0.00000  &       & 0.21859  &       & 0.21740  &       & 0.21706  &       & 0.21741  &       & 0.41061  &       & 0.39707  \\
				&       &       &       & \textbf{meanT} &       & 0.00000  &       & 0.31149  &       & 0.00658  &       & 0.00557  &       & 0.00660  &       & 0.00625  &       & 0.00418  \\
				\bottomrule
			\end{tabular}%
		\end{threeparttable}
	}
	\caption{ATT - Details on Balancing Metrics with 500 Monte Carlo datasets.}
	\label{Table4ATTBalanceMetric_X}%
\end{table}%

\subsubsection{Simulation 2}\label{subsubsec:sim2}
In this section, we perform a simulation experiment with a parameter setting different from Simulation 1 (section \ref{subsubsec:sim1}). We wish to explore the mechanism of the KDB method in mining covariate information as well as check the performance of the KDB stable ATE estimation. 

We first construct the treatment indicator $T_i, ~i =1,2,\cdots,N$ with a binomial distribution $ Bin(1,p) $. Here, we set $ p $, the probability for unit $ i $ being assigned to the treatment group, to be 0.5. 
Next, we used two different multivariate normal distributions to generate the covariates $ X_1 $ and $ X_2 $ for the treatment group and the control group, respectively. We have 
   \[
   \begin{pmatrix}
   X_1 &X_2
   \end{pmatrix}|T= 1 \sim N\begin{pmatrix}
   \begin{pmatrix}
   1 \\ 2
   \end{pmatrix} , \begin{pmatrix}
   1 & 0.5\\ 0.5 & 1
   \end{pmatrix}
   \end{pmatrix}
   \]
and 
   \[
   \begin{pmatrix}
   X_1 &X_2
   \end{pmatrix}|T= 0 \sim N\begin{pmatrix}
   \begin{pmatrix}
   1 \\ 2
   \end{pmatrix} + \alpha_1 , \begin{pmatrix}
   1 & 0.5\\ 0.5 & 1
   \end{pmatrix} + \begin{pmatrix}
   0 & \alpha_2 \\ \alpha_2 & 0
   \end{pmatrix}
   \end{pmatrix},
   \]
where $\alpha_1 = 0.8,~\alpha_2 = 0.2$. Let $ X_3 $ and $ X_4 $ to be distributer as $ \mathcal{N}(0,1) $, we continue to construct the covariates $ X_5 = X_1 \times X_2 $ and $ X_6 = X_3^2 $. The vector $ (X_1, X_2, X_3, X_4, X_5, X_6)^T $ is subsequently standardized to have a mean of zero and marginal variances of one. The observed covariates are $ \{X_1, X_2, X_3, X_4\} $. For the outcome variable, we use the bivariate model
   \[
   \begin{pmatrix}
   Y_i(0)\\Y_i(1)
   \end{pmatrix} \sim N\begin{pmatrix}
   \left[ \begin{array}{c}
   \mu_i\\\mu_i+\gamma
   \end{array} 
   \right ] , \left[ \begin{array}{cc}
   \sigma^2 & 0 \\
   0 & \sigma^2\\
   \end{array} 
   \right] 
   \end{pmatrix}
   \]
where $\gamma = 10 $, $ \sigma^2 =10$, and $ \mu_i = 20X_{i1}+10X_{i2}+5X_{i3}+5X_{i4} + \alpha_3 \cdot X_{i5} + \alpha_4 \cdot X_{i6}  $ with $\alpha_3 = 1, \alpha_4 = 2$.
To explore whether KDB stable ATE estimation can effectively reduce the variance of ATE estimator, we add a tuning parameter $\lambda$ to the diagonal of $ \bm{K}_G $ and change the value to see see how the estimation result changes.
   \[
   \bm{K_G}  = \begin{pmatrix}
   K_1 & -K_{10} \\
   -K_{01} & K_0 \\
   \end{pmatrix} +\lambda \bm{I}_N
   \] 
Setting the sample size $ N = 200 $, we run 500 Monte Carlo datasets for each value of $ \lambda $ in $ \{0,1,2,5,10,100\} $. Results of ATE estimation with $ \lambda \in  \{0,1,2,5,10,100\} $ are shown in Table [\ref{ATETable4Simulation4_lambda6val}]. In each table for a specific value of $ \lambda $, the rows are arranged from top to bottom according to the absolute value of bias.

Comparing the results in all tables together, we can find that the UnAD estimator always performs the worst, and is far from the true value. This shows the importance of covariate balance, especially when the real generation mechanism of the data involves nonlinear information. 
As the value of $ \lambda $ increases, the performance of KDBC gradually deteriorates, but the performance of KDM1 has always been in the forefront, which indicates the necessity to add the first-moment balance to the balance constraints of the KDB method. Furthermore, adding a non-zero $ \lambda $ do help reduce the variance of the KDM1 estimator of ATE compared to the results when $ \lambda = 0 $, the mechanism behind which requires to be further explored. 

When $ \lambda $ changes from 0 to 1, the performance of KDM1 is not as good as CBPS and CAL. However, when we increase the $ \lambda $ value from 2 to 100, KDM1 again becomes the best weighting method with the smallest absolute value of bias, the mechanism behind which needs further exploration. This probably implies that there is a threshold for $ \lambda $ to induce better performance, and it may be related to the information matrix (section \ref{subsec:Gkd}) of the observed dataset.

Considering the ASMD values of $X_5$ and $X_6$ under different weighting methods in Table [\ref{ATETable4Simulation4_lambda6val}], we know that the KDB method including the KDBC and KDM1 can always perform the best. 
Reviewing the performance of the KDB covariate balance in simulation 1 (section \ref{subsubsec:sim1}) with \{$ \delta_T $ = X, $ \delta_O $  = U\} and \{$ \delta_T $ = U, $ \delta_O $  = U\}, we can conclude that the KDB covariate balance may automatically discover and include some latent variables, such as interaction terms between different covariates or high-order terms of covariates, to analyze and estimate the ATE.

\begin{table}[H]
	\centering
	\resizebox{\linewidth}{!}{
		\begin{threeparttable}
		\begin{tabular}{cccccccccccccc}
			&       &       &       &       &       &       &       &       &       &       &       &       &  \\
			\multicolumn{14}{c}{ lambda = 0 } \\
			\midrule
			& \textbf{ATE} & \textbf{abs(Bias)} & \textbf{sd} & \textbf{RMSE} & \textbf{rw} & \textbf{KD} & \textbf{maxASMD} & \textbf{meanASMD} & \textbf{medASMD} & \textbf{X5ASMD} & \textbf{X6ASMD} & \textbf{meanKS} & \textbf{meanT} \\
			\midrule
			\textbf{Oracle} & 9.98058  & 0.01942  & 0.01427  & 0.01428  & 0.00000  & 0.00000  & 0.00000  & 0.00000  & 0.00000  & 0.00000  & 0.00000  & 0.00000  & 0.00000  \\
			\textbf{KDM1} & 9.97253  & 0.02747  & 0.02993  & 0.02993  & 0.00000  & 0.00034  & 0.00179  & 0.00039  & 0.00000  & 0.00068  & 0.00167  & 0.23125  & -0.01172  \\
			\textbf{KDBC} & 9.95925  & 0.04075  & 0.02982  & 0.02985  & 0.00000  & 0.00034  & 0.00203  & 0.00078  & 0.00061  & 0.00128  & 0.00166  & 0.23229  & -0.01277  \\
			\textbf{CBPS} & 9.93864  & 0.06136  & 0.02546  & 0.02559  & 0.00087  & 0.02829  & 0.15146  & 0.03478  & 0.00001  & 0.08006  & 0.12859  & 0.23308  & -0.00203  \\
			\textbf{CAL} & 9.92957  & 0.07043  & 0.02471  & 0.02488  & 0.00063  & 0.02436  & 0.13794  & 0.03100  & 0.00000  & 0.06410  & 0.12190  & 0.23103  & -0.00062  \\
			\textbf{IPW} & 9.87181  & 0.12819  & 0.13501  & 0.13500  & 0.00155  & 0.03605  & 0.18029  & 0.07931  & 0.06737  & 0.11296  & 0.13136  & 0.23359  & -0.05327  \\
			\textbf{EB} & 9.86523  & 0.13477  & 0.02452  & 0.02522  & 0.00063  & 0.02435  & 0.13861  & 0.03212  & 0.00218  & 0.06553  & 0.12184  & 0.23103  & -0.00725  \\
			\textbf{UnAD} & -13.16629  & 23.16629  & 0.15097  & 1.04695  & 0.04264  & 0.20455  & 0.89249  & 0.46910  & 0.46869  & 0.87363  & 0.11004  & 0.24557  & -2.83703  \\
			\midrule
			&       &       &       &       &       &       &       &       &       &       &       &       &  \\
			\midrule
			\midrule
			&       &       &       &       &       &       &       &       &       &       &       &       &  \\
			\multicolumn{14}{c}{lambda = 1} \\
			\midrule
			& \textbf{ATE} & \textbf{abs(Bias)} & \textbf{sd} & \textbf{RMSE} & \textbf{rw} & \textbf{KD} & \textbf{maxASMD} & \textbf{meanASMD} & \textbf{medASMD} & \textbf{X5ASMD} & \textbf{X6ASMD} & \textbf{meanKS} & \textbf{meanT} \\
			\midrule
			\textbf{Oracle} & 9.98058  & 0.01942  & 0.01427  & 0.01428  & 0.00000  & NA    & 0.00000  & 0.00000  & 0.00000  & 0.00000  & 0.00000  & 0.00000  & 0.00000  \\
			\textbf{CBPS} & 9.93864  & 0.06136  & 0.02546  & 0.02559  & 0.00087  & 0.02829  & 0.15146  & 0.03478  & 0.00001  & 0.08006  & 0.12859  & 0.23308  & -0.00203  \\
			\textbf{CAL} & 9.92957  & 0.07043  & 0.02471  & 0.02488  & 0.00063  & 0.02436  & 0.13794  & 0.03100  & 0.00000  & 0.06410  & 0.12190  & 0.23103  & -0.00062  \\
			\textbf{KDM1} & 9.92614  & 0.07386  & 0.02230  & 0.02252  & 0.00027  & 0.01609  & 0.09301  & 0.02067  & 0.00000  & 0.04032  & 0.08370  & 0.22262  & -0.02399  \\
			\textbf{IPW} & 9.87181  & 0.12819  & 0.13501  & 0.13500  & 0.00155  & 0.03605  & 0.18029  & 0.07931  & 0.06737  & 0.11296  & 0.13136  & 0.23359  & -0.05327  \\
			\textbf{EB} & 9.86523  & 0.13477  & 0.02452  & 0.02522  & 0.00063  & 0.02435  & 0.13861  & 0.03212  & 0.00218  & 0.06553  & 0.12184  & 0.23103  & -0.00725  \\
			\textbf{KDBC} & 7.70684  & 2.29316  & 0.03258  & 0.10759  & 0.00068  & 0.02577  & 0.13520  & 0.06471  & 0.06194  & 0.11691  & 0.08285  & 0.22935  & -0.29548  \\
			\textbf{UnAD} & -13.16629  & 23.16629  & 0.15097  & 1.04695  & 0.04264  & 0.20455  & 0.89249  & 0.46910  & 0.46869  & 0.87363  & 0.11004  & 0.24557  & -2.83703  \\
			\midrule
			&       &       &       &       &       &       &       &       &       &       &       &       &  \\
			\midrule
			\midrule
			&       &       &       &       &       &       &       &       &       &       &       &       &  \\
			\multicolumn{14}{c}{lambda = 2} \\
			\midrule
			& \textbf{ATE} & \textbf{abs(Bias)} & \textbf{sd} & \textbf{RMSE} & \textbf{rw} & \textbf{KD} & \textbf{maxASMD} & \textbf{meanASMD} & \textbf{medASMD} & \textbf{X5ASMD} & \textbf{X6ASMD} & \textbf{meanKS} & \textbf{meanT} \\
			\midrule
			\textbf{Oracle} & 9.99897  & 0.00103  & 0.01403  & 0.01402  & 0.00000  & 0.00000  & 0.00000  & 0.00000  & 0.00000  & 0.00000  & 0.00000  & 0.00000  & 0.00000  \\
			\textbf{CAL} & 9.97598  & 0.02402  & 0.02712  & 0.02711  & 0.00061  & 0.02399  & 0.14136  & 0.03183  & 0.00000  & 0.06461  & 0.12634  & 0.23060  & 0.00053  \\
			\textbf{KDM1} & 9.97495  & 0.02505  & 0.02545  & 0.02544  & 0.00034  & 0.01796  & 0.11197  & 0.02460  & 0.00000  & 0.04486  & 0.10274  & 0.22157  & -0.02585  \\
			\textbf{CBPS} & 9.96688  & 0.03312  & 0.02809  & 0.02810  & 0.00085  & 0.02795  & 0.15344  & 0.03519  & 0.00001  & 0.08018  & 0.13097  & 0.23332  & -0.00194  \\
			\textbf{EB} & 9.91031  & 0.08969  & 0.02739  & 0.02766  & 0.00061  & 0.02397  & 0.14194  & 0.03297  & 0.00220  & 0.06612  & 0.12629  & 0.23060  & -0.00639  \\
			\textbf{IPW} & 9.86231  & 0.13769  & 0.14389  & 0.14388  & 0.00165  & 0.03577  & 0.18388  & 0.07980  & 0.06765  & 0.11501  & 0.13453  & 0.23366  & -0.06557  \\
			\textbf{KDBC} & 5.90692  & 4.09308  & 0.04408  & 0.18827  & 0.00172  & 0.04099  & 0.20144  & 0.10423  & 0.10464  & 0.18551  & 0.09985  & 0.23222  & -0.52996  \\
			\textbf{UnAD} & -13.21379  & 23.21379  & 0.15533  & 1.04968  & 0.04286  & 0.20487  & 0.89554  & 0.47282  & 0.47129  & 0.87486  & 0.11403  & 0.24615  & -2.86241  \\
			\midrule
			&       &       &       &       &       &       &       &       &       &       &       &       &  \\
			\midrule
			\midrule
			&       &       &       &       &       &       &       &       &       &       &       &       &  \\
			\multicolumn{14}{c}{lambda = 5} \\
			\midrule
			& \textbf{ATE} & \textbf{abs(Bias)} & \textbf{sd} & \textbf{RMSE} & \textbf{rw} & \textbf{KD} & \textbf{maxASMD} & \textbf{meanASMD} & \textbf{medASMD} & \textbf{X5ASMD} & \textbf{X6ASMD} & \textbf{meanKS} & \textbf{meanT} \\
			\textbf{Oracle} & 9.99204  & 0.00796  & 0.01455  & 0.01454  & 0.00000  & 0.00000  & 0.00000  & 0.00000  & 0.00000  & 0.00000  & 0.00000  & 0.00000  & 0.00000  \\
			\textbf{KDM1} & 9.97729  & 0.02271  & 0.02752  & 0.02751  & 0.00042  & 0.02001  & 0.12573  & 0.02743  & 0.00000  & 0.04826  & 0.11631  & 0.22127  & -0.02422  \\
			\textbf{CAL} & 9.97347  & 0.02653  & 0.02839  & 0.02839  & 0.00063  & 0.02423  & 0.14342  & 0.03196  & 0.00000  & 0.06393  & 0.12784  & 0.23043  & -0.00077  \\
			\textbf{CBPS} & 9.96713  & 0.03287  & 0.02915  & 0.02916  & 0.00085  & 0.02797  & 0.15523  & 0.03541  & 0.00001  & 0.07816  & 0.13429  & 0.23273  & -0.00133  \\
			\textbf{EB} & 9.90745  & 0.09255  & 0.02870  & 0.02897  & 0.00063  & 0.02421  & 0.14408  & 0.03316  & 0.00223  & 0.06564  & 0.12778  & 0.23045  & -0.00754  \\
			\textbf{IPW} & 9.71187  & 0.28813  & 0.12863  & 0.12914  & 0.00154  & 0.03547  & 0.18570  & 0.07895  & 0.06512  & 0.11247  & 0.13634  & 0.23305  & -0.06307  \\
			\textbf{KDBC} & 1.99376  & 8.00624  & 0.07158  & 0.36512  & 0.00566  & 0.07427  & 0.34591  & 0.18240  & 0.18200  & 0.33050  & 0.10952  & 0.23495  & -1.02101  \\
			\textbf{UnAD} & -13.08489  & 23.08489  & 0.15413  & 1.04381  & 0.04235  & 0.20387  & 0.89282  & 0.47006  & 0.46973  & 0.87439  & 0.11324  & 0.24486  & -2.82364  \\
			\midrule
			&       &       &       &       &       &       &       &       &       &       &       &       &  \\
			\midrule
			\midrule
			&       &       &       &       &       &       &       &       &       &       &       &       &  \\
			\multicolumn{14}{c}{lambda = 10} \\
			\midrule
			& \textbf{ATE} & \textbf{abs(Bias)} & \textbf{sd} & \textbf{RMSE} & \textbf{rw} & \textbf{KD} & \textbf{maxASMD} & \textbf{meanASMD} & \textbf{medASMD} & \textbf{X5ASMD} & \textbf{X6ASMD} & \textbf{meanKS} & \textbf{meanT} \\
			\midrule
			\textbf{Oracle} & 10.00065  & 0.00065  & 0.01403  & 0.01402  & 0.00000  & 0.00000  & 0.00000  & 0.00000  & 0.00000  & 0.00000  & 0.00000  & 0.00000  & 0.00000  \\
			\textbf{KDM1} & 9.98450  & 0.01550  & 0.02582  & 0.02581  & 0.00044  & 0.02045  & 0.12810  & 0.02795  & 0.00000  & 0.04786  & 0.11985  & 0.22178  & -0.02839  \\
			\textbf{CAL} & 9.97525  & 0.02475  & 0.02652  & 0.02652  & 0.00060  & 0.02373  & 0.13972  & 0.03153  & 0.00000  & 0.06203  & 0.12714  & 0.23110  & -0.00074  \\
			\textbf{CBPS} & 9.97272  & 0.02728  & 0.02745  & 0.02745  & 0.00081  & 0.02729  & 0.14996  & 0.03443  & 0.00001  & 0.07602  & 0.13054  & 0.23393  & -0.00384  \\
			\textbf{EB} & 9.91211  & 0.08789  & 0.02633  & 0.02660  & 0.00060  & 0.02370  & 0.14022  & 0.03266  & 0.00214  & 0.06360  & 0.12710  & 0.23110  & -0.00735  \\
			\textbf{IPW} & 9.57129  & 0.42871  & 0.11484  & 0.11632  & 0.00140  & 0.03458  & 0.17781  & 0.07669  & 0.06409  & 0.11294  & 0.13090  & 0.23390  & -0.07975  \\
			\textbf{KDBC} & -1.93687  & 11.93687  & 0.09537  & 0.54227  & 0.01188  & 0.10779  & 0.48651  & 0.25751  & 0.25467  & 0.47040  & 0.11033  & 0.23875  & -1.53728  \\
			\textbf{UnAD} & -13.28347  & 23.28347  & 0.15303  & 1.05243  & 0.04300  & 0.20539  & 0.89664  & 0.47257  & 0.47019  & 0.87847  & 0.11167  & 0.24624  & -2.87076  \\
			\midrule
			&       &       &       &       &       &       &       &       &       &       &       &       &  \\
			\midrule
			\midrule
			&       &       &       &       &       &       &       &       &       &       &       &       &  \\
			\multicolumn{14}{c}{lambda = 100} \\
			\midrule
			& \textbf{ATE} & \textbf{abs(Bias)} & \textbf{sd} & \textbf{RMSE} & \textbf{rw} & \textbf{KD} & \textbf{maxASMD} & \textbf{meanASMD} & \textbf{medASMD} & \textbf{X5ASMD} & \textbf{X6ASMD} & \textbf{meanKS} & \textbf{meanT} \\
			\midrule
			\textbf{Oracle} & 9.99204  & 0.00796  & 0.01455  & 0.01454  & 0.00000  & NA    & 0.00000  & 0.00000  & 0.00000  & 0.00000  & 0.00000  & 0.00000  & 0.00000  \\
			\textbf{KDM1} & 9.97934  & 0.02066  & 0.02821  & 0.02820  & 0.00050  & 0.02168  & 0.13688  & 0.02977  & 0.00000  & 0.05193  & 0.12669  & 0.21941  & -0.02443  \\
			\textbf{CAL} & 9.97347  & 0.02653  & 0.02839  & 0.02839  & 0.00063  & 0.02423  & 0.14342  & 0.03196  & 0.00000  & 0.06393  & 0.12784  & 0.23043  & -0.00077  \\
			\textbf{CBPS} & 9.96713  & 0.03287  & 0.02915  & 0.02916  & 0.00085  & 0.02797  & 0.15523  & 0.03541  & 0.00001  & 0.07816  & 0.13429  & 0.23273  & -0.00133  \\
			\textbf{EB} & 9.90745  & 0.09255  & 0.02870  & 0.02897  & 0.00063  & 0.02421  & 0.14408  & 0.03316  & 0.00223  & 0.06564  & 0.12778  & 0.23045  & -0.00754  \\
			\textbf{IPW} & 9.71187  & 0.28813  & 0.12863  & 0.12914  & 0.00154  & 0.03547  & 0.18570  & 0.07895  & 0.06512  & 0.11247  & 0.13634  & 0.23305  & -0.06307  \\
			\textbf{KDBC} & -11.07394  & 21.07394  & 0.14527  & 0.95356  & 0.03548  & 0.18653  & 0.82059  & 0.43223  & 0.43112  & 0.80211  & 0.11307  & 0.24343  & -2.61005  \\
			\textbf{UnAD} & -13.08489  & 23.08489  & 0.15413  & 1.04381  & 0.04235  & 0.20387  & 0.89282  & 0.47006  & 0.46973  & 0.87439  & 0.11324  & 0.24486  & -2.82364  \\			
			\bottomrule
		\end{tabular}%
		\end{threeparttable}
	}
	\caption{ $\lambda \in \{0,1,2,5,10,100\}$.}
	\label{ATETable4Simulation4_lambda6val}
\end{table}%

\subsection{Real Data Analysis}\label{subsec:rda}
To further compare different weighting methods, we investigate the causal effect of a real data set - the labour training program data that previously analyzed in \citet{lalonde1986evaluating} and \citet{dehejia1999causal}, among many others.

The dataset used here is a subset containing information on earnings in 1974 (RE74) of the National Supported Work Demonstration as used in \citet{lalonde1986evaluating}. The National Supported Work Demonstration is an evaluation of the National Supported Work Demonstration project, a transitional, subsidized work experience program for four target groups of people with longstanding employment problems: ex-offenders, former drug addicts, women who were long-term recipients of welfare benefits, and school dropouts, many with criminal records. It is designed to test whether and to what extent 12 to 18 months of employment in a supportive but performance-oriented environment would equip hard-to-employ people to get and hold regular, unsubsidized jobs.

\begin{table}[H]
	\centering
	\resizebox{\linewidth}{!}{
		\begin{threeparttable}
			\begin{tabular}{ccccccccccccccccc}
				\toprule
				\textbf{Metric} &       & \textbf{UnAD} &       & \textbf{IPW} &       & \textbf{CBPS} &       & \textbf{EB} &       & \textbf{CAL} &       & \textbf{KDBC} &       & \textbf{KDps} &       & \textbf{KDM1} \\
				\midrule
				ATE &       & 1804.78064  &       & 1641.77770  &       & 1637.56504  &       & 1612.77550  &       & 1612.44289  &       & 1774.39211  &       & 2311.99732  &       & 1672.08084  \\
				\textbf{sd} &       & 28.78274  &       & 29.73068  &       & 29.73643  &       & 30.01748  &       & 30.01800  &       & 29.48966  &       & 56.44856  &       & 29.32907  \\
				\midrule
				\midrule
				\textbf{Balancing  Metric} &       & \textbf{UnAD} &       & \textbf{IPW} &       & \textbf{CBPS} &       & \textbf{EB} &       & \textbf{CAL} &       & \textbf{KDBC} &       & \textbf{KDps} &       & \textbf{KDM1} \\
				\midrule
				\textbf{KD} &       & 0.00051  &       & 0.00012  &       & 0.00004  &       & 0.00004  &       & 0.00004  &       & 0.00002  &       & 0.00002  &       & 0.00001  \\
				\textbf{maxASMD} &       & 0.00862  &       & 0.00802  &       & 0.00801  &       & 0.00800  &       & 0.00800  &       & 0.00864  &       & 0.00928  &       & 0.00803  \\
				\textbf{meanASMD} &       & 0.00480  &       & 0.00445  &       & 0.00445  &       & 0.00445  &       & 0.00445  &       & 0.00482  &       & 0.00476  &       & 0.00445  \\
				\textbf{medASMD} &       & 0.00498  &       & 0.00460  &       & 0.00460  &       & 0.00460  &       & 0.00460  &       & 0.00499  &       & 0.00469  &       & 0.00464  \\
				\textbf{meanKS} &       & 0.57085  &       & 0.42960  &       & 0.42476  &       & 0.40371  &       & 0.40311  &       & 0.54584  &       & 0.22214  &       & 0.43027  \\
				\textbf{meanT} &       & 9.81470  &       & 6.94619  &       & 6.86728  &       & 6.79432  &       & 6.78668  &       & 8.98585  &       & 2.07824  &       & 7.34122  \\
				\bottomrule
			\end{tabular}%
		\end{threeparttable}
	}
	\caption{Results of ATE for the NSW data \citep{dehejia1999causal}. The ATE is an average of 500 bootstrap estimates, and $ \bm{sd = \frac{sd(\bm{ATE})}{\sqrt{500}}} $ is the standard error of the average ATE estimates.}
	\label{Table4NSW}%
\end{table}%

Variables included in the dataset are: treatment indicator (1 if treated, 0 if not treated), age, education, Black (1 if black, 0 otherwise), Hispanic (1 if Hispanic, 0 otherwise), married (1 if married, 0 otherwise), nodegree (1 if no degree, 0 otherwise), RE74 (earnings in 1974), RE75 (earnings in 1975), and RE78 (earnings in 1978). We compared the weighting estimates for earnings in 1978 between the treatment and control groups, which were estimates for the treatment effects. 
We compare the proposed KDB covariate balance with the IPW, CBPS, EB, CAL, and take the UnAD estimator as the benchmark. 
We set the balance restrictions in the KDB method as follows: regularization constraint only, regularization constraints plus propensity score balance, and regularization constraints plus the first-moment balance. Each scenario is denoted by ``KDBC'', ``KDps'', and ``KDM1'', respectively. We obtain 500 bootstrap datasets using repeatable sampling, with which we estimate the ATE. The results are shown in Table [\ref{Table4NSW}], and all numerical results are rounded to five decimal places. 
As can be seen from Table [\ref{Table4NSW}], the estimation results of all weighting methods are significantly different from those of UnAD estimates. Except for the KDB method (KDBC, KDps, KDM1), the estimation results of other weighting methods are very similar. 
Both KDBC and KDM1 have the smallest standard deviation (sd), so they show outstanding stability in estimating ATE. Considering the specific estimates of ATE, however, only adding regularization constraints may not be enough to achieve a good balance of covariates between the treatment and control group. 
In addition, the results of KDPS estimation are completely different from those of other methods including UnAD, and its sd is very large. Therefore, it seems inappropriate to use regularization constraints plus propensity score balance as balance restrictions in the KDB method. This may be an important hint to remind us to carefully set balance restrictions, which is an important exploration focus.

\begin{figure}[H]
	\centering
	\includegraphics[width=0.8\textheight,height=0.95\textheight]{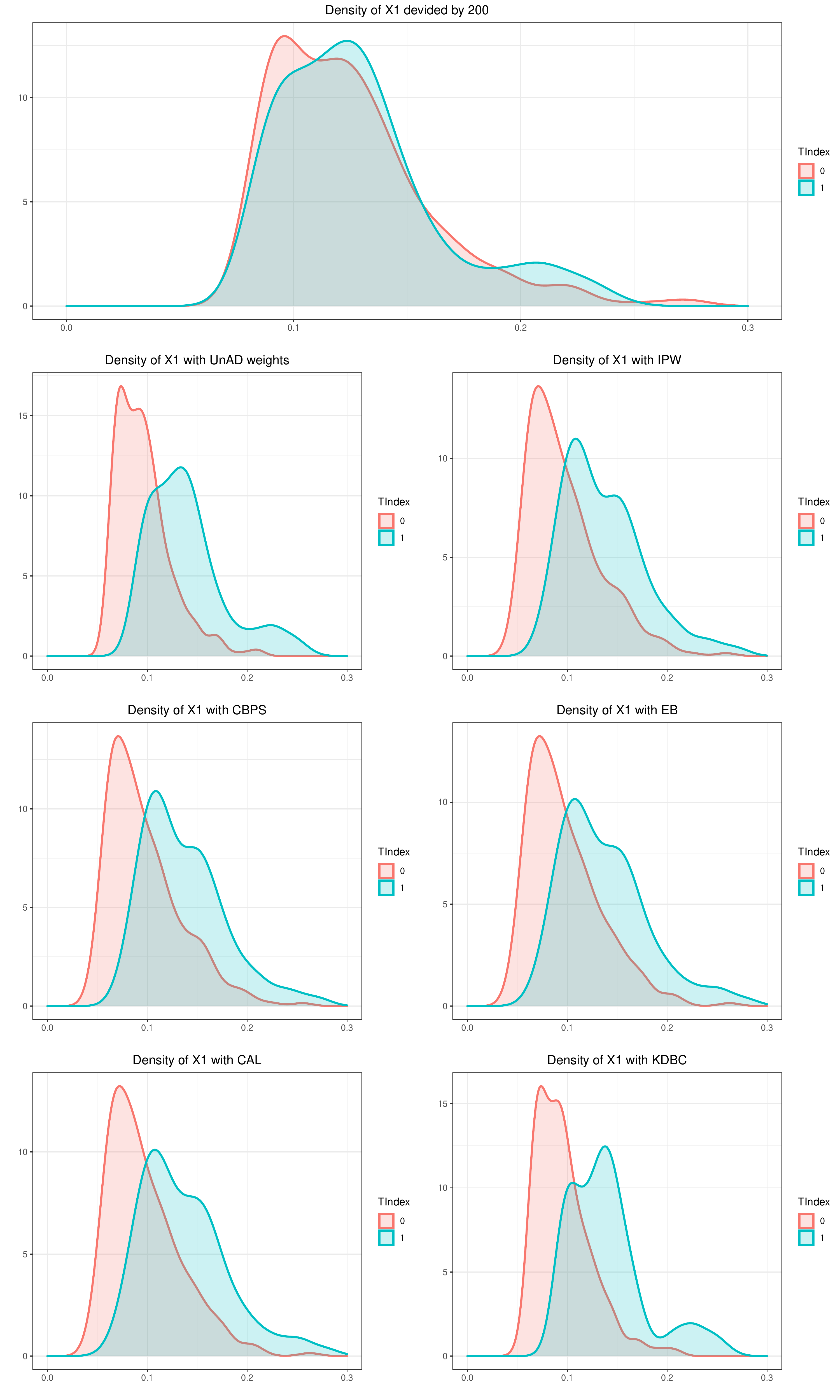}
	\caption{Density of $X_1 = age$ under different balancing methods.}
	\label{RDA_PDF_X1}
\end{figure}

\begin{figure}[H]
	\centering
	\includegraphics[width=0.8\textheight,height=0.95\textheight]{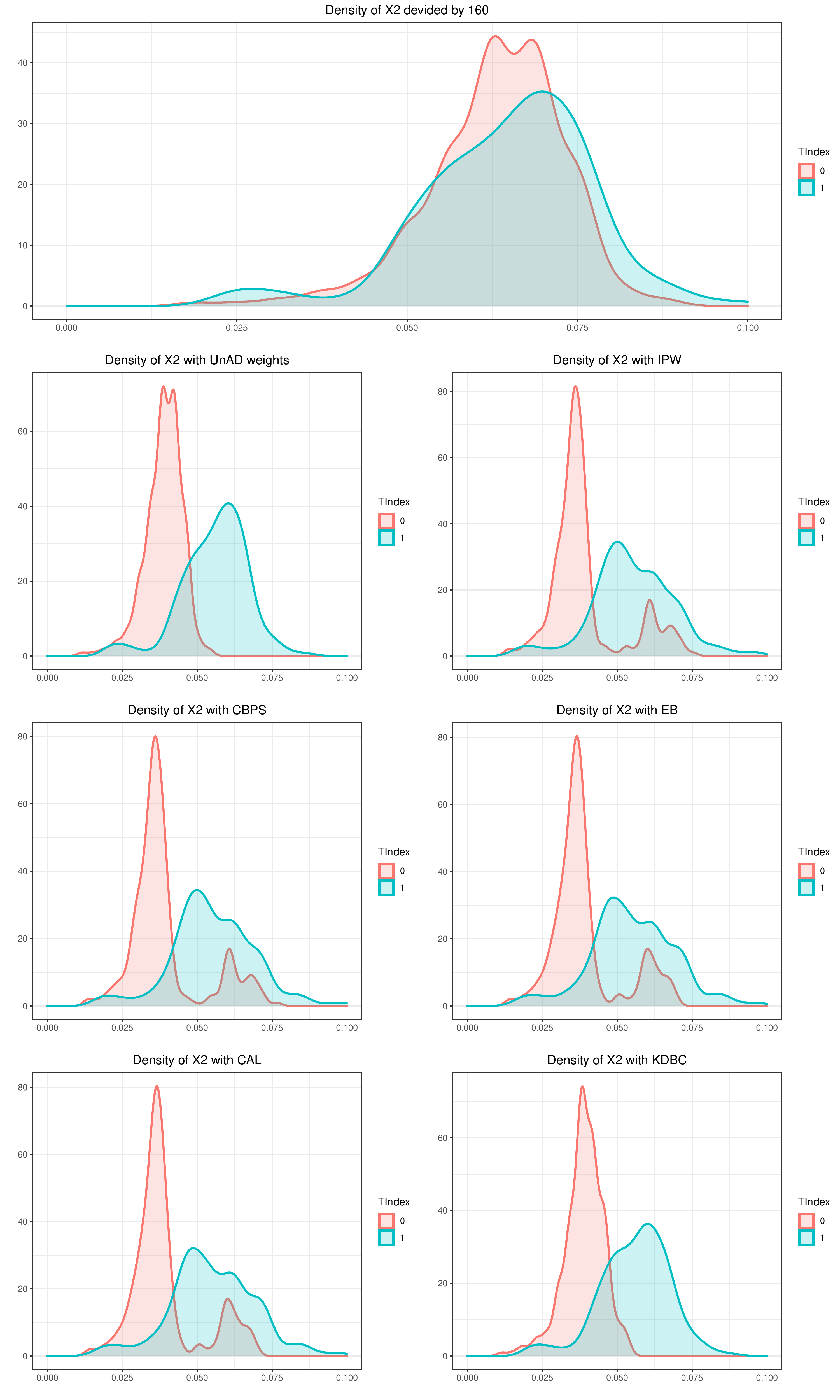}
	\caption{Density of $X_2 = education$ under different balancing methods.}
	\label{RDA_PDF_X2}
\end{figure}

We plot the probability density functions (PDF) of covariates ``age'' (denoted by $ X_1 $) and ``education'' (denoted by $ X_2 $), as well as PDFs of $ X_1 $  and $ X_2 $ weighted by different methods (UnAD, IPW, CBPS, EB, CAL, KDBC) to compare performance. Both the first PDF plot in Figure [\ref{RDA_PDF_X1}] and Figure [\ref{RDA_PDF_X2}] indicate that $ X_1 $ and $ X_2 $ have almost the same distributions in the treatment and control groups, which is consistent with the randomization setting in NSW data. 
In Figure [\ref{RDA_PDF_X1},\ref{RDA_PDF_X2}], the relative relationship between the PDFs of $ X_1 $ and $ X_2 $ weighted by the IPW, CBPS, EB, and CAL methods in the treatment and control groups is very similar, which may be able to explain why the estimation results of these weighting methods in Table [\ref{Table4NSW}] are highly similar.
Although all methods have widened the gap between the probability density functions of the treatment group and the control group, the KDBC method seems to better retain the characteristics of the original PDF, such as bimodal.

\newpage

\newpage

\section{Conclusions and Further Discussions}\label{sec:conNdis}
\subsection{Conclusions}\label{subsec:con}
With the goal of creating balanced samples, we proposed a kernel-distance-based covariate balancing method with the Gaussian kernel in this article. The proposed method works for calculating sample weights for the treatment and control groups in order to yield proper estimator for the causal effect. It seems to perform well in discovering the latent variable, such as the interaction of different predictors, and yield stable estimate for the ATE.

\subsection{Further Discussions}\label{subsec:dis}
There are two main directions for our proposed method to develop. (i) In addition to setting the kernel distance as the loss function, we can also choose other commonly used covariate balance metrics as the loss function, and use the kernel distance as an optimization constraint, so that it is possible to control the balance performance from different dimensions at the same time. (ii) We look forward to exploring the further applications of kernel distance in high-dimensional space.



\section*{Acknowledgments}

\nocite{*}
\bibliographystyle{ecta}
\bibliography{MyRef}


\end{document}